\documentclass[useAMS, usenatbib]{mnras}
\usepackage{amsmath}
\usepackage{verbatim}
\usepackage{graphicx}
\usepackage{url}
\usepackage{setspace}
\usepackage{longtable}
\usepackage{pdflscape}
\usepackage{float}
\usepackage{booktabs}
\usepackage{tabularx}
\usepackage{amssymb }
\usepackage{xcolor}

\usepackage[T1]{fontenc}

\batchmode

\newcommand{\kms}{km s$^{-1}$}

\makeatletter
\def\fps@figure{htbp}
\makeatother

\makeatletter
\def\fps@table{htbp}
\makeatother

\title[Disc-Halo Decompositions in Spiral Galaxies ]{The Influence of a Kinematically Cold Young Component on Disc-Halo Decompositions
in Spiral Galaxies: Insights from Solar Neighbourhood K-giants}
\author[S. Aniyan et al.]{S. Aniyan$^{1}$\thanks{Email: suryashree.aniyan@anu.edu.au}, K. C. Freeman$^1$, O. E. Gerhard$^2$, M. Arnaboldi$^{3,4}$ \& C. Flynn$^5$ \\
$^1$Research School of Astronomy \& Astrophysics, Australian National University, via Cotter Road, Weston, ACT 2611, Australia \\
$^2$Max-Planck-Institut f{\"u}r Extraterrestrische Physik, Giessenbachstrasse, 85741 Garching, Germany\\
 $^3$European Southern Observatory, Karl-Schwarzschild Str. 2, 85748 Garching Germany \\
$^4$INAF, Osservatorio Astronomico di Torino, Strada Osservatorio 20, 10025 Pino Torinese, Italy\\
$^5$Centre for Astrophysics and Supercomputing, Swinburne University, Hawthorn, VIC 3122, Australia}
\begin{document}

\maketitle
\begin{abstract}
In decomposing the HI rotation curves of disc galaxies, it is necessary to break a degeneracy between the gravitational fields of the disc and the dark halo by estimating the disc surface density. This is done by combining measurements of the vertical velocity dispersion of the disc with the disc scale height. The vertical velocity dispersion of the discs is measured from absorption lines (near the V-band) of near-face-on spiral galaxies, with the light coming from a mixed population of giants of all ages. However, the scale heights for these galaxies are estimated statistically from near-IR surface photometry of edge-on galaxies. The scale height estimate is therefore dominated by a population of older ($> 2$ Gyr) red giants.

In this paper, we demonstrate the importance of measuring the velocity dispersion for the same older population of stars that is used to estimate the vertical scale height. We present an analysis of the vertical kinematics of K-giants in the solar vicinity. We find the vertical velocity distribution best fit by two components with dispersions of 9.6 $\pm$ 0.5 km s$^{-1}$ and 18.6 $\pm$ 1.0 km s$^{-1}$, which we interpret as the dispersions of the young and old disc populations respectively. Combining the (single) measured velocity dispersion of the total young + old disc population (13.0 $\pm$ 0.1 km s$^{-1}$) with the scale height estimated for the older population would underestimate the disc surface density by a factor of $\sim 2$. Such a disc would have a peak rotational velocity that is only 70\% of that for the maximal disc, thus making it appear submaximal.

\end{abstract}


\section{Introduction}
Galactic rotation curves are currently the best way to measure parameters for dark haloes of spiral galaxies, such as their typical scale densities and scale lengths. These quantities are interesting in themselves but they are also cosmologically significant, because the densities and scale radii of dark haloes follow well-defined scaling laws which can be used to measure the redshift of assembly of haloes of different masses \citep{Kormendy14, Maccio:13}.

The density profiles of the dark haloes of spiral galaxies can be measured by decomposing their 21-cm rotation curves into the contributions from the disc and the dark halo. In the decomposition process, the shape of the disc's contribution to the rotation curve is calculated from the radial light distribution of the disc. It is then scaled according to the adopted mass-to-light ratio (M/L) of the disc. For the solar neighnourhood, the M/L values for the solar neighbourhood can be determined directly from star counts: e.g. \citet{Just:15} for the near IR and \citet{Flynn:06} for optical bands. They find M/L $\sim$ 1.5, 1.2 and 0.34 (M/L)$_\odot$ in the V, I and K-band respectively. These values cannot be universally applied to other Galactic regions and other discs because M/L depends on the local star formation history. The stellar M/L values for external discs are still uncertain. As the adopted M/L of the disc is increased, the disc contributes increasingly more to the rotation curve, and the resulting halo becomes less dense and has a longer scale length. The adopted M/L for the disc is critical to the outcome. The M/L can be estimated from the rotation curve itself, but this suffers from the well-known degeneracy between the contributions from the disc and the dark halo \citep{vanAlbada}. 

The stellar M/L can in principle be estimated from stellar population synthesis (SPS) models (e.g. \citealt{Bell:2001}) but this remains insecure as it requires several significant assumptions such as the star formation and chemical enrichment history, the stellar initial mass function (IMF), accurate accounting of late phases of stellar evolution (e.g. \citealt{Maraston05}) and the internal dust absorption \citep{Tully:1985}.

To break the disc-halo degeneracy, the vertical velocity dispersion of stars in the discs can be used to measure the surface mass density of the disc (e.g. \citealt{Bottema97}, \citealt{PNI}). From the 1D Jeans equation in the vertical direction, the vertical velocity dispersion $\sigma_z$ (integrated vertically through the disc) and the vertical disc scale height $h$ together give the surface mass density $\Sigma$ of the disc from the simple relation:\\
\begin{equation}
 \Sigma = f\sigma_z^2/Gh
\end{equation}
 where G is the gravitational constant and $f$ is a geometric factor that depends weakly on the adopted vertical structure of the disc. The surface brightness of the disc and the surface mass density ($\Sigma$ from Eqn. 1) together give the M/L of the disc. The factor $f = 2/3\pi$ for a vertically exponential disc, and $f = 1/2\pi$ for a vertically isothermal disc \citep{vdKF2011}. The velocity dispersion $\sigma_z$ is measured from spectra of the integrated light of the disc in relatively face-on galaxies. These observations are difficult because high resolution spectra of low surface brightness discs are required to measure the small dispersions (for the old disc near the sun, $\sigma_z \sim$ 20 km s$^{-1}$). The other parameter, the disc scale height $h$, is typically about 250 pc, but cannot be measured directly for these face-on galaxies. It has to be estimated statistically from similar galaxies, seen edge-on, using the relation between scale height and absolute magnitude or circular velocity that has been measured for samples of edge-on galaxies. \citet{YandD} show the correlation of the scale heights of the thin and thick disc with circular velocity of edge-on disc galaxies using R-band data. Similarly, \citet{Kregel} did I-band studies of edge-on disc galaxies and show correlations between the scale height and intrinsic properties of the galaxy such as its surface brightness. \citet{VanFree}, \citet{Bottema} and \citet{DMI} have used this method and find that the disc M/L is relatively low and the discs are submaximal \footnote{A maximal disc has the maximum M/L value consistent with the observed rotation curve and a non-hollow dark halo. Typically, the disc of a maximal disc provides about 85\% of the rotational velocity at the peak of the rotation curve \citep{Sackett:97}.}. \citet{DM} find that the dynamical stellar M/L obtained is about 3 times lower than the M/L from maximum disc hypothesis.
 
Equation (1) comes from the vertical Jeans equation for an equilibrium disc. It is therefore essential that the vertical disc scale height, $h$, and the vertical velocity dispersion, $\sigma_z$, should refer to the same population of stars. This raises a potential problem.

The dispersion $\sigma_z$ is usually measured from integrated light spectra near the Mg b lines ($\sim 5150 - 5200$ \AA), since this region has many absorption lines and the sky is relatively dark. The discs of the gas-rich galaxies for which good HI rotation data are available usually have a continuing history of star formation and therefore include a population of young (ages $< 2$ Gyr), kinematically cold stars among a population of older, kinematically hotter stars, as we will demonstrate in this paper for the Galactic disc near the sun. The red giants of this mixed young + old population provide most of the absorption line signal that is used for deriving velocity dispersions from the integrated light spectra of galactic discs. These same giants typically contribute about half of the light of the V-band integrated light spectra of discs (see Appendix). On the other hand, red and near-infrared measurements of the scale heights of edge-on disc galaxies are dominated by the red giants of the older, kinematically hotter population (see bottom panel in Fig. 1). The dust layer near the Galactic plane further weights the determination of the scale height to the older kinematically hotter population: e.g. \citet{de-Grijs:97}.

Therefore, in Eqn. 1, we should be using the velocity dispersion of the older disc stars in combination with the scale heights of this same population for an accurate determination of the surface mass density. In practice, because of limited signal-to-noise ratios for the integrated light spectra of the discs, integrated light measurements of the disc velocity dispersions usually adopt a single kinematical population for the velocity dispersion whereas, ideally, the dispersion of the older stars should be extracted from the composite observed spectrum of the younger and older stars. 

Adopting a single kinematical population for a composite kinematical population gives a velocity dispersion that is smaller than the velocity dispersion of the old disc giants (for which the scale height was measured), and hence underestimates the surface density of the disc. A maximal disc will then appear submaximal. This problem potentially affects the usual dynamical tracers of the disc surface density in external galaxies, like red giants and planetary nebulae, which have progenitors covering a wide range of ages. It therefore affects most of the previous studies. \citet{Flynn:94} demonstrated the presence of bright, kinematically colder stars in the solar neighbourhood. They recognised the importance of isolating the old K-giants for dynamical measures of the surface density of the Galactic disc.

Fig. 1 presents a synthetic colour magnitude diagram computed using IAC-Star (see \citealt{IAC}) with the Teramo stellar evolution library and the Castelli \& Kurucz bolometric correction library, to show the location of the younger and older red giants. The adopted star formation history has the exponential form $\exp(-t/\beta)$ with $\beta$ = 20 Gyr. We used a Kroupa IMF and a linear chemical enrichment law, with a mean metallicity Z = 0.006 and 0.019 at t = 0 Gyr and t = 13 Gyr respectively. We introduced a spread of $\pm\, 0.4$ dex in the metallicities at all ages, to match the dispersion in the observed age-metallicity law in the solar neighborhood (e.g. \citealt{Haywood08}). Stars younger than $2$ Gyr are shown in red in Fig. 1 and the black points are for older stars. Fig. 1 shows that the younger giants are more likely to be among the most luminous stars on the giant branch (see Appendix for the fraction of total light contributed by the giants). We also show the $M_K, J-K$ colour-magnitude diagram for the same simulation, to indicate the contribution that the older giants make to the near-IR surface brightness in photometric studies of edge-on discs.
\begin{figure} 
\includegraphics[width=0.48\textwidth]{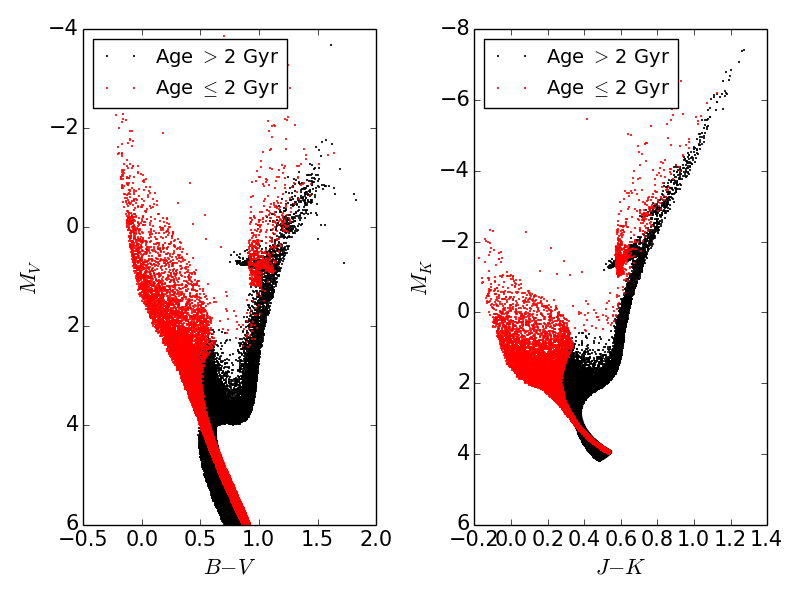}
\caption{The CMD computed from IAC-STAR for a disc with an exponentially declining star formation law. The red points are stars with ages $\leq$ 2 Gyr. The left panel shows the CMD in the V-band. The right panel shows the CMD in the near-infrared. The old giants are the brightest stars in IR.}
\end{figure}

\subsection{The surface density of the Galactic disc near the sun}

So far, we have discussed only the use of integrated light spectroscopy and estimates of the disc scale height to measure the surface density of the discs of external galaxies. In the Milky Way, the velocities and spatial distributions of individual tracer stars
near the sun can be used for the same purpose. We briefly review the results in Table 1, because they provide a useful context for the integrated light observations of external galaxies. 

\begin{table} 
\resizebox{0.48\textwidth}{!}{%
\begin{tabular}{llcl}
\hline
Author &$\Sigma_{disc}$ & $\Sigma_{total}(1.1$ kpc) & Sample \\ \hline
\citet{KGI, KGII, KGIII} & $48 \pm 8$ M$_\odot$ pc$^{-2}$ & $71 \pm 6$ M$_\odot$ pc$^{-2}$ & K-dwarfs \\ 
\citet{Flynn:94} & $52 \pm 13$ M$_\odot$ pc$^{-2}$ & & K-giants \\ 
\citet{Holmberg2004} & $56 \pm 6$ M$_{\odot}$ pc$^{-2}$ & $74 \pm 6$ M$_\odot$ pc$^{-2}$ & K-giants\\ 
\citet{BR} & $38 \pm 4$ M$_\odot$ pc$^{-2}$  &$68 \pm 4$ M$_\odot$ pc$^{-2}$ &G-dwarfs\\ \hline
\end{tabular}
}
\caption{Studies that have looked at the solar neighbourhood to estimate the surface density of the Galactic disc and the total surface density (disc + dark halo).}
\end{table}

The values obtained from these different studies agree well. The implications regarding the maximality of the Milky
Way's disc are uncertain, because the radial scale length of the Galactic disc is not accurately known. 

\subsection{The Disk Mass Survey}
\citet{DMI} used integral field spectroscopy to study the kinematics of stars and gas in the discs of face-on spiral galaxies out to radii of about 2.2 scalelengths. The discs in their sample contribute typically 15\% to 30\% of the dynamical mass within 2.2 disc scalelengths, with percentages increasing systematically with luminosity, rotation speed, and redder color. These trends indicate that the mass ratio of disc-to-total matter remains at or below 50\% at 2.2 scale length, even for the most rapidly rotating discs ($V_{max} \geq 300$ km s$^{-1}$). The conclusion is that spiral discs are generally submaximal \citep{DM}. As in all previous dynamical studies of this kind, these authors model the stars in the disc as a single kinematical population and determine a single vertical velocity dispersion to represent all of the stars.

\subsection{Are discs really so submaximal?}
In the following sections, we analyse the K-giants in the solar neighbourhood, to demonstrate the existence of different populations of stars with different vertical velocity dispersions and discuss the implications for the decomposition of the HI rotation curves of external galaxies. While the K-giants dominate the velocity dispersion signal in the integrated spectra of external discs, it is only near the Sun that we can examine in detail the composite kinematical makeup of the K-giant population, and work out the implications for integrated light spectroscopy of an external disc that has a similar star formation history to the Galactic disc near the Sun. In particular, we would like to evaluate the significance of a cold core of K-giants among the kinematically hotter stars of the old disc. 

Section 2 describes our data sample and section 3 describes the analysis. The results are discussed in section 4. Section 5 uses the Besan\c{c}on model as a check on any unanticipated selection effects that our sample of giants may have, section 6 uses the Besan\c{c}on model to look at the implications for external galaxies, section 7 has our conclusions and section 8 discusses our future plans for this project. 

\section{Selection of Sample}
We would like to construct a sample of nearby red giants that has age and kinematics for each star. Currently, these data are not yet available but are soon expected from asteroseismology of red giants (e.g. \citealt{Soderblom2010}). At present, guided by Fig. 1, we can use giants of known vertical velocity (W velocity) and absolute magnitude to look at their distribution over kinematics and luminosity. 
We use two different samples to build up our data set -- a sample of K giants in the South Galactic Pole (SGP), for which the W velocities come mainly from the radial velocities, and giants from the Bright Star Catalog. 

The \citet{ff} sample (hereafter FF) is a sample of 560 K-giants at the SGP with $V < 11.0$. It was selected from the following sources:
\begin{itemize}
 \item Henry Draper (HD) stars with spectral types between G8 and K5, brighter than luminosity class V, from the Michigan Catalog Volumes III and IV \citep{houk, houk2},
 \item The \citeauthor{Eriksson95} (1978/1995) sample of SGP stars with $0.95 < B-V < 1.55$ and $V < 11.0$, 
 \item A sample of fainter giants from the Zeiss 6-inch camera on the Oddie telescope at Mount Stromlo Observatory (see FF).
\end{itemize}

We used a subset of 303 stars from the FF sample with measured $B - V$, radial velocity and absolute magnitude ($M_V$). The absolute magnitudes in the FF sample were originally estimated using the intermediate-band photometric David Dunlop Observatory (DDO) system. \citet{Holmberg2004} later compared the DDO absolute magnitudes with the more accurate Hipparcos absolute magnitudes and found some systematic offsets in the DDO system. The absolute magnitudes of our stars from the FF sample were re-calibrated as suggested by \citet{Holmberg2004}. The expected absolute magnitude of the red giant branch clump stars ($M_V$$\sim$ 0.8) is consistent with the revised magnitude scale.

The sample of 303 FF red giant stars with colour, magnitude and radial velocity information is not large enough to evaluate the detailed structure of the W velocity distribution function, so we increased our sample size by adding giants from the Bright Star Catalog (BSC) \citep{BSC}. The BSC is more or less complete to $V = 7$.
We chose stars from the BSC that have the same spectral type as the FF stars, i.e from G8III to K5III.
 
\section{Analysis}
In this section, we calculate the space velocities for our sample of stars and represent the distribution of W velocity in terms of different populations of stars.
\subsection{Vertical Velocities of the Sample of Giants}
Of the above subset of 303 stars from the FF sample, proper motions for 300 are available from the UCAC4 catalogue \citep{UCAC4}; 134 of them have parallaxes and associated errors from the extended Hipparcos catalogue \citep{XHIP}. Photometric distances for all 300 stars were calculated using the data in FF and the corrected $M_V$ values using the method described by \citet{Holmberg2004}. The $M_V$ values have errors of 0.35 mag \citep{Holmberg2004} and the apparent V magnitudes in \citet{ff} have typical errors $\sim 0.02$ mag. Therefore, our photometric distances typically have a 16\% error. For the final calculation of W velocities, we compared the relative errors of the data from the Hipparcos catalog (where available) and the photometric distances, and used the distance with the smaller relative error. 

The BSC catalog has radial velocities with typical errors \textless \ 1 km s$^{-1}$. It also contains the parallaxes and proper motion from the Hipparcos catalog. Our final sample now contains 1740 stars.

The stellar W velocities were calculated as described in \citet{JS}, with the W velocity positive towards the North Galactic Pole. The errors of the W velocities were again calculated as in \citet{JS}. For our study of the W velocity distribution function, only stars with W velocity errors \textless \ 5 km s$^{-1}$ were retained in the sample to avoid contamination of the velocity distribution by measurement errors. Large errors will compromise our attempt to recover the young component that has small dispersions. Our sample now contains W velocities for 1567 stars. Fig. 2 shows their W velocities against the absolute magnitude, $M_V$. The W velocities are heliocentric, and the solar motion (our data has mean $W = -7$ km s$^{-1}$) is evident in Fig. 2. The rapid decrease in the stellar density for stars with $M_V > 1$ results from the apparent magnitude and spectral type limits of our sample. The smaller velocity dispersion of the younger disc giants ($M_V$ \textless\ 0) is also evident (cf. Fig. 1, upper right panel). The older stars show a visibly larger spread in W velocities. Fig. 3 shows the $\sigma_W$ vs. $M_V$ for our sample. The low dispersion of the colder component at $M_V < -2$ is clearly visible. \citet{Flynn:94} also show the presence of this cold, bright population of stars in the solar neighbourhood.

\begin{figure} 
\includegraphics[width=0.5\textwidth]{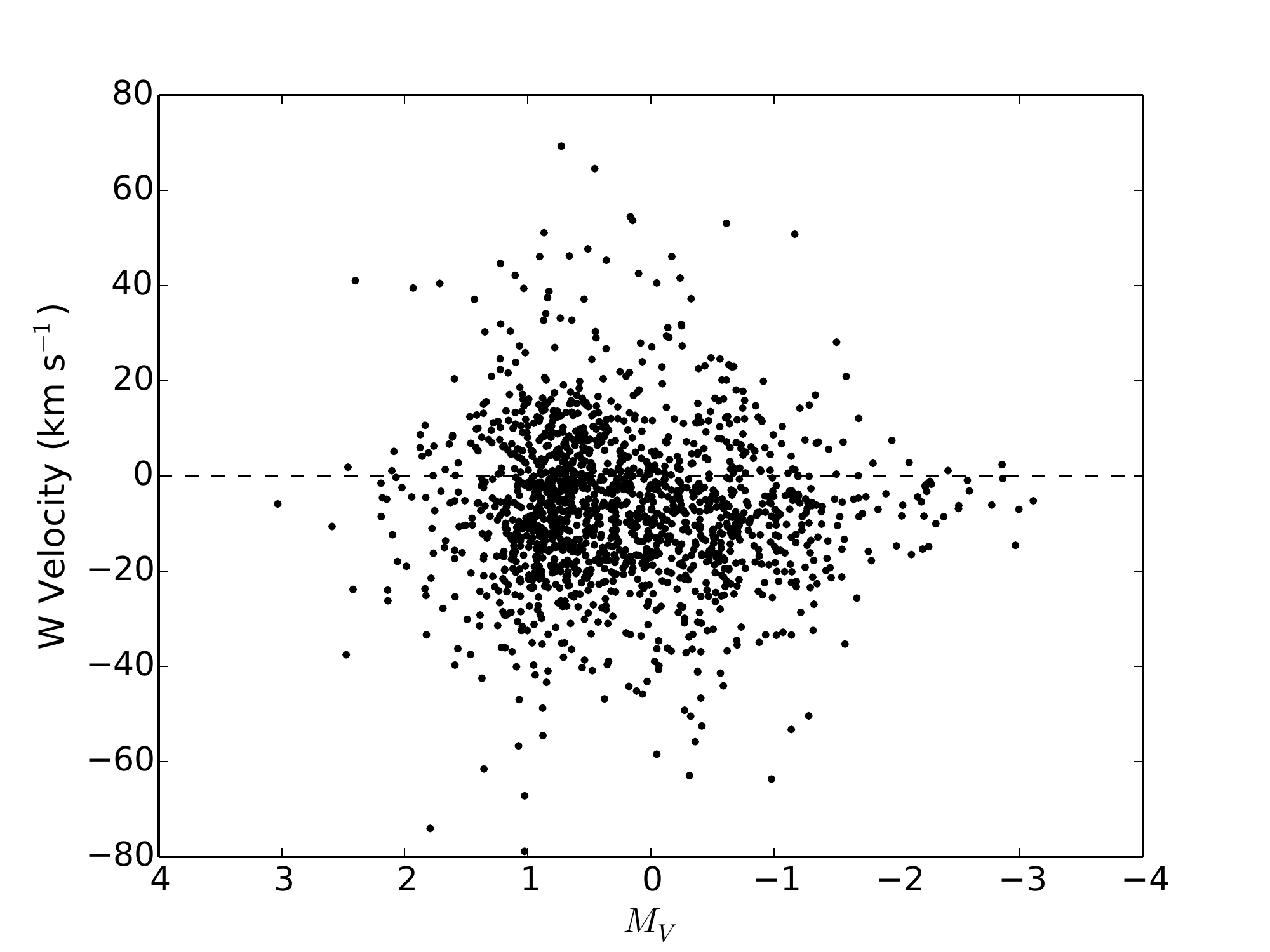}
\caption{W velocities versus absolute magnitude for the nearby giants. The colder stars with smaller dispersions can been seen at $M_V \leq -2$. The offset from zero of the mean W-velocity is due to 7 km s$^{-1}$ reflex W-motion of the sun.}
\end{figure}

\begin{figure} 
\includegraphics[width=0.5\textwidth]{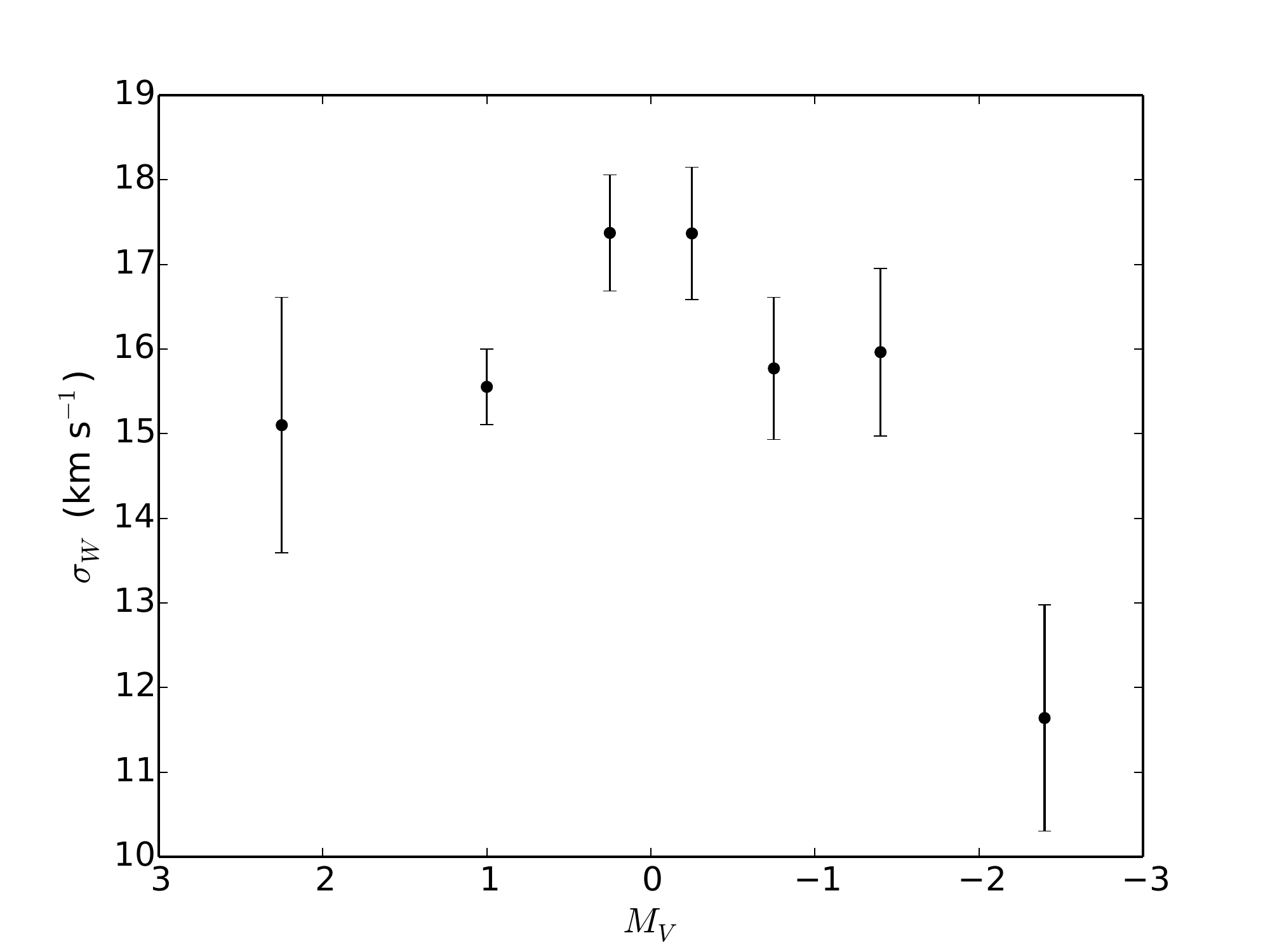}
\caption{The vertical velocity dispersion ($\sigma_W$) versus absolute magnitude for the nearby giants. The brighter stars ($M_V < -2$) are much colder as indicated by their low $\sigma_W$ value.}
\end{figure}

\subsection{Age-Velocity Relation}
It has long been known that the velocity dispersion of stars in the Galactic thin disc near the Sun depends on their age (see e.g. \citealt{Delhaye65} for a summary). The young disc stars with ages \textless\ 2 Gyr have $\sigma_W \sim$ 10 - 12 \kms\ and stars older than $\sim 2$ Gyr have $\sigma_W \sim$ 20 \kms. Several more recent works have looked at the structure of the age-velocity relation of thin disc stars and have found similar results \citep{Freeman91, Edvardsson93, Gomez97, Quillen00}: a steady increase in $\sigma_W$ for stellar ages up to $\sim 2-3$ Gyr and then a roughly constant velocity dispersion for stars with ages between about $3 - 10$ Gyr. Based on these results, if we were to look at the W velocity distribution of thin disc stars near the sun, we would find that the data could be well represented by two Gaussians -- one with a standard deviation of $\sim 10-12$ \kms\ representing the younger stars and the other with a standard deviation of $\sim$ 20 \kms\ representing the older stars.

Other studies \citep[e.g.][]{Wielen77} indicate that the velocity dispersion of the thin disc stars does not plateau for the older stars but continues to increase steadily with age. In their study of the Geneva-Copenhagen sample, \citet{Casagrande11} find that the velocity dispersion continues to increase up to an age of $\sim 10$ Gyr. Their derived rate of increase for the dispersion of the older stars depends on the adopted stellar models and the abundance and age cuts imposed on the sample (see their Fig. 17). If this continuously rising velocity dispersion with age represents the W velocity distribution of the nearby thin disc, then a simple two-component distribution would not be a good representation of the velocity distribution. A more complex model would be required. For the purpose of this paper, we will adopt the former model, assuming that
two approximately Gaussian velocity components (older and younger) are present among the giants of the thin disc.

\subsection{Multiple Populations of Stars}
Our aim is to determine whether the kinematical parameters for two different population of stars (i.e the younger, colder population and the older, hotter component) can be extracted from our sample of W velocities of nearby stars. In order to do this, we constructed a generalized histogram of W velocities by representing the velocity of each star as a Gaussian of unit area with mean at its observed W velocity and the error in W velocity as its standard deviation (see Fig. 4). The motivation for using a generalized histogram was to avoid the effects of binning. Our Gaussian models are of the form:

\begin{equation}
\sum_{k=1}^{n} A_k e^{-(x - \mu_k)^2/2\sigma_k^2}
\end{equation}

\noindent where A is the amplitude, $\mu$ is the mean and $\sigma$ is the standard deviation of the Gaussian. For a single Gaussian model, $n = 1$ and for a double Gaussian model, $n = 2$.

We used the curve\_fit module in the python scipy.optimize package to determine the parameters for the two Gaussians. Curve\_fit uses non-linear least squares and the Levenberg-Marquardt algorithm. It requires an initial guess of parameters and returns the best-fit values, along with the covariance matrix. The errors associated with each parameter are the square roots of the corresponding elements on the diagonal of the covariance matrix. To fit a two-Gaussian model, we need to estimate 6 parameters i.e the amplitude, mean and standard deviation for each component. We will also be fitting single Gaussian models (3 parameters) for comparison. 

In order to evaluate whether a two-component model is preferred over a one-component model, we compute the
Akaike Information Criterion ($AIC$) which can be used to determine whether adding additional parameters results
in a better fit. The $AIC$ is defined by

\begin{equation}
AIC = n\ln(SSE) - n\ln(n) + 2p
\end{equation}

\noindent where $n$ is the number of samples, $p$ is the number of parameters in the model, and $SSE$ is the sum of the squared errors 

\begin{equation}
SSE = \Sigma (y_{data} - y_{model})^2.
\end{equation} 

The AIC penalizes additional parameters in the model; the model with the lowest AIC is preferred.

\section{THE TWO VELOCITY COMPONENTS}
We now derive the double Gaussian and single Gaussian fits to the distribution of W velocities for our sample of K-giants. A significant cold component is visible in the double Gaussian fit. We show again that the brightest giants ($M_V \leq -1.8$ ) are kinematically very cold, while the fainter giants are an almost homogenous mixture of the hot and cold populations.

Fig. 4 shows the generalized velocity histogram of our sample of stars (black), fit with a two component model
(red). It demonstrates the presence of a colder population of stars among a hotter disc population. 

\begin{figure} 
\includegraphics[width=0.5\textwidth]{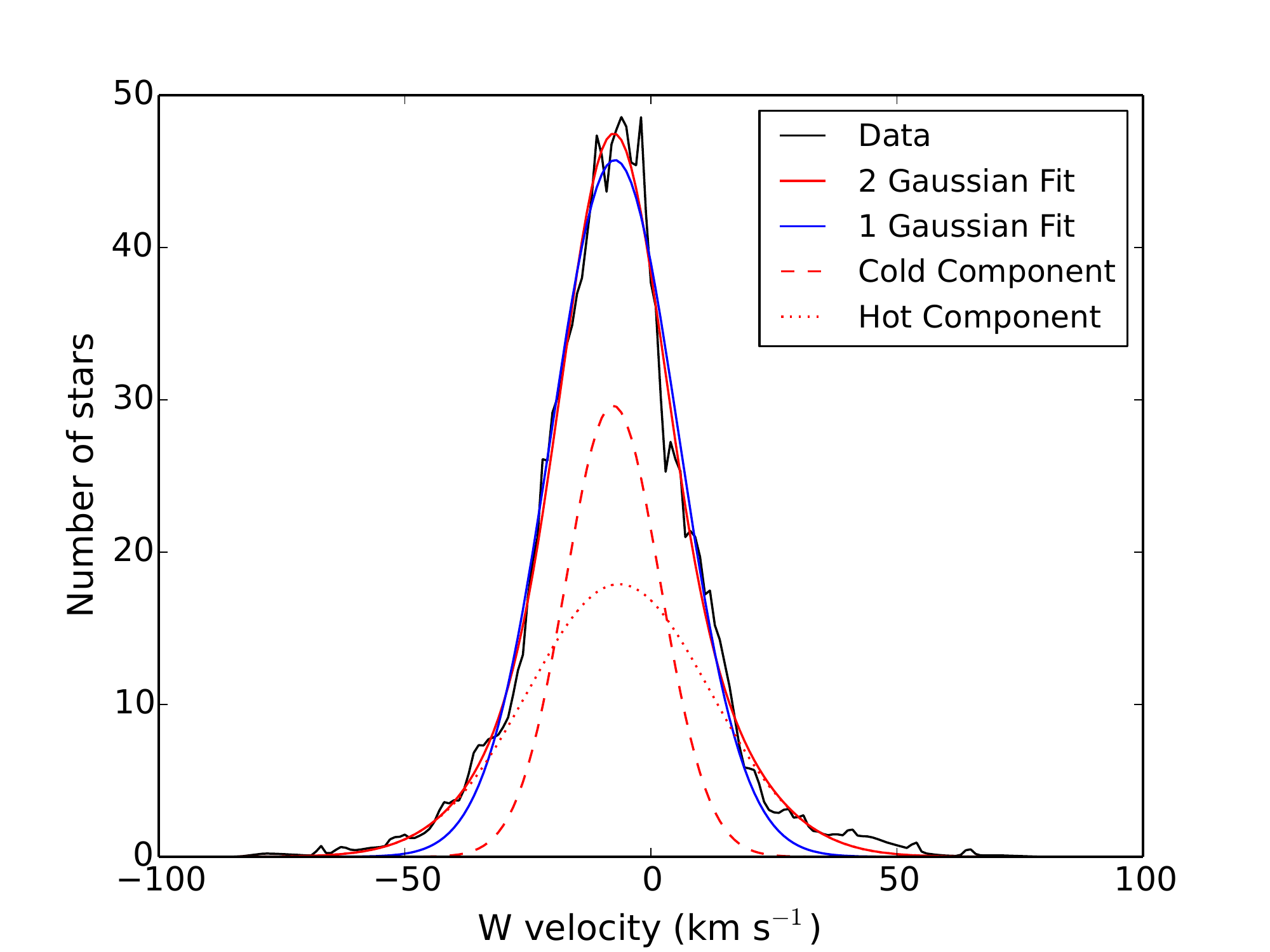}
\caption{Generalised histogram of W velocities for the combined red giants from \protect\cite{ff} and BSC (in black). The red curve is the best fit to the data, using 2 Gaussians to represent the colder (red dashed line) and hotter (red dotted line) components of the thin disc. The blue curve represents the best single-Gaussian fit, for comparison.}
\end{figure}

The results of our single Gaussian fit and double Gaussian fit are tabulated in Table 2. 

\begin{table} 
\resizebox{0.48\textwidth}{!}{%
\begin{tabular}{cccc}
\hline
&Cold Component & Hot Component & Single Gaussian Fit \\ \hline
Amplitude & $29.6 \pm 3.1$ & $17.9 \pm 3.2$ & $45.7 \pm 0.4$ \\ 
Mean (km s$^{-1}$) & $-7.7 \pm 0.2$ & $-6.5 \pm 0.5$ & $-7.3 \pm 0.1$ \\ 
Sigma (km s$^{-1}$) & $9.6 \pm 0.5$ & $18.6 \pm 1.0$ & $13.0 \pm 0.1$ \\ 
AIC & \multicolumn{2}{c}{$55.2$} & $155.7$ \\ \hline
\end{tabular}
}
\caption{The fit results for the 1 component and 2 component models. The lower value of the AIC indicates that the two Gaussian model is preferred, despite the larger number of parameters.}
\end{table}

Using Eqn. 3, we get an AIC = 55.2 for the 2 component model and AIC = 155.7 for the 1-component model. Clearly, the 2-component model is preferred in this case. Fig. 4 shows the fit to the generalized histogram using two-component and one-component fits. The 2-component fit is visibly better than a single Gaussian fit. 

The goal is to use the measured velocity dispersion and the scale height to calculate the surface density of galactic discs that are viewed more or less face-on. The adopted velocity dispersion and the calculated surface density depend on the kinematical model (one component or two), and we now estimate how much difference this can make. We use Eqn. 1 which relates the surface density $\Sigma$, the scale height $h$ and the velocity dispersion $\sigma$ of the disc.  The scale height cannot be measured directly: it is estimated from the scaling laws for the scale height, and pertains to the old disc population, as explained in section 1.  The underlying dynamical assumption is that the disc is in vertical equilibrium in its own gravitational potential, so the scale height and the velocity dispersion should be for the same population. What is the effect on the surface density of using a single-component kinematical model to represent the disc, when in reality there are two kinematical components?  The true velocity dispersion of the old disc is then the dispersion $\sigma_{\rm hot}$ of the hotter of the two components, and the true surface density of the disc is $\Sigma_{\rm true} = f \sigma_{\rm hot}^2/Gh$. On the other hand, if we were to use a single component kinematical model, we would calculate its surface density as 
$\Sigma_{\rm single} = f \sigma_{\rm single}^2/Gh$. The velocity dispersion $\sigma_{\rm single}$ for the single-component model is lower than $\sigma_{\rm hot}$, as in Table 2 ($13.0$ km s$^{-1}$ and $18.6$ km s$^{-1}$ resp.). The adopted scale height $h$ is not affected by the kinematical measurements, so the effect of using the one-component model is to underestimate the surface density of the disc.

Taking, in each case, the same scale height $h$  for the old thin disc, the ratio of surface densities derived from using the dispersion for (a) the hotter component of a double Gaussian fit and (b) a single Gaussian fit, is
\begin{equation}
\frac{\Sigma_{2}}{\Sigma_{1}} = \frac{\sigma_2^2}{\sigma_1^2} = \frac{(18.6 \pm 1.0)^2}{(13.0 \pm 0.1 )^2} = 2.05 \pm 0.16
\end{equation}

\noindent The subscript 1 refers to the case where we fit a single Gaussian and subscript 2 refers to the hotter (older) component of the double Gaussian fit. If we were using integrated light spectroscopy to measure the surface density of the disc of an external galaxy, in which the star formation history over the last 10 Gyr has been similar to that in the solar neighbourhood, and if we had used a single-component Gaussian velocity distribution instead of deriving the velocity dispersion of the old disc from a two-component velocity distribution, then we would underestimate the surface density of the disc by about a factor 2.

We can use Eqn. 1 and the dispersion for the hotter component of the two-component model in Table 2, to estimate the surface density of the disc near the sun and compare with the more detailed estimates of surface density given in Table 1. We assume that the old disc is vertically exponential, with a scale height $h =  300 \pm 50$ pc (e.g. \citealt{Gilmore:83}) and a velocity  dispersion of $\sigma = 18.6 \pm 1.0$ km s$^{-1}$ as given in Table 2. The surface density $\Sigma = (2/3\pi)\,\sigma^2/Gh = 57 \pm 12$ M$_\odot$ pc$^{-2}$, in fair agreement with the range of estimates given for the surface density of the disc in Table 1.

In this section, we have shown that the W velocity distribution of the nearby K-giants is well represented by a
two-component distribution. Fig. 3 shows that the more luminous giants have a smaller W velocity dispersion than the fainter giants, and we associate the two components with the older and younger stars of the thin disc. 

We can visualise and quantify this effect further, by splitting the sample of stars into 3 absolute magnitude intervals: $M_V \leq -1.8, -1.8 <$ $M_V \leq 0$ and $M_V > 0$. Within each group, the stars are ordered by increasing absolute magnitude. We then form the cumulative sum of $\mid W - \overline{W}\mid$ vs the rank of the star, where 
$\overline{W}$ is the mean value of $W$ for the group. The point of doing this kind of analysis is that, for a group of stars with homogeneous velocity dispersion (not necessarily a single Gaussian component), the plots of the sum of $\mid W - \overline{W}\mid$ against rank will be a straight line, and the slope is a measure of the velocity dispersion. Fig. 5 shows the outcome for each of the three absolute magnitude groups.

\begin{figure} 
\includegraphics[width=0.49\textwidth]{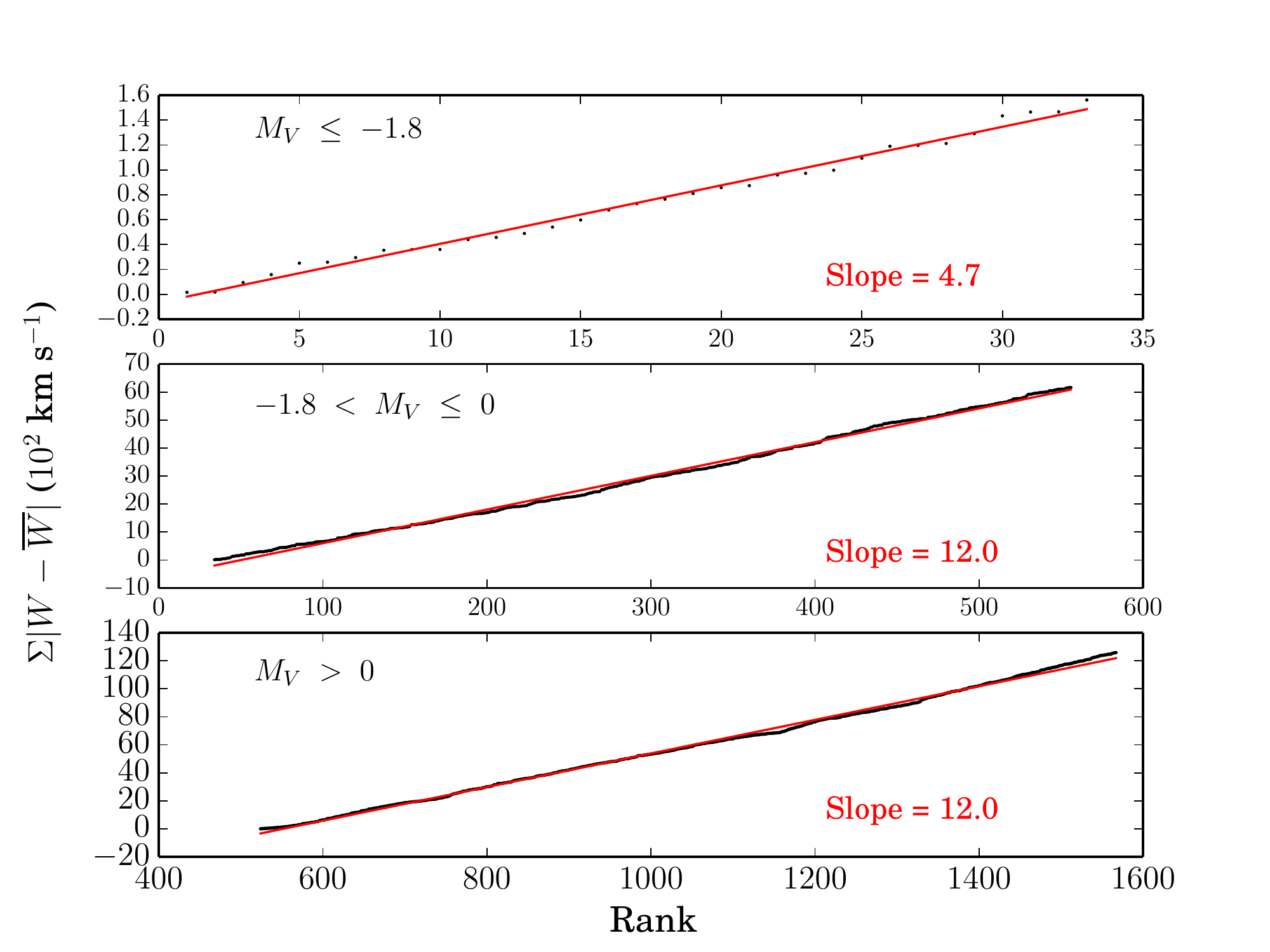}
\caption{The cumulative sum of $\mid W - \overline{W}\mid$ against rank by absolute magnitude in three magnitude intervals for our giants. The x-axis shows the rank of the stars and the y-axis is the cumulative sum of W velocities. See section 4 in the text for further details. The upper panel is for the bright, young, kinematically colder population. The stars in the two bottom panels contain a mixed population of young and old stars. The velocity dispersion of the population is proportional to the slope of the line.}
\end{figure}

The red line in each panel of Fig. 5 is a linear fit to the data. The slope of the line $= \sqrt{2/\pi}\,\,\sigma$, where $\sigma$ is the velocity dispersion. Fitting the slope of the line is equivalent to deriving a single dispersion for the velocity distribution of the stars, after setting their mean velocity to zero. The bright giants with $M_V \leq -1.8$ represent mainly the younger and kinematically colder population. Their dispersion is $\sigma = 5.9$ \kms\, which is colder than the value we obtained for the young, cold component in our two-component Gaussian decomposition. 

The two panels of fainter stars with $M_V > -1.8$ are mixed populations of stars of all ages, with similar (larger) dispersions. While the brighter stars in the top panel are clearly from a colder population, we conclude from Fig. 5 that the mixtures of hot and cold populations in the two lower panels are fairly similar and do not change much with absolute magnitude, because the two panels are well fit by straight lines and have similar slopes with dispersions of 15 km s$^{-1}$.

\section{Comparison with kinematics of the Besan\c{c}on model}
As a check on what we have learned about the velocity distribution of the nearby thin disc from our sample of nearby giants, we did a similar study on a sample of simulated giants from the on-line Besan\c{c}on model \citep{Bes}. This model was designed to represent the main Galactic structural components and stellar populations. It includes recipes for galactic reddening, the star formation history and the dynamical evolution which manifests as the stellar age-velocity dispersion relation \citep[Table 4]{Bes}. Our main goal in using the Besan\c{c}on model here is as a check on any significant unanticipated selection effects in our sample of giants.

Two separate simulated catalogs were generated -- one to represent the FF stars and another for the BSC stars. The simulation that mimics the FF giant sample was made by choosing all stars within a distance of 1.5 kpc, $-3 \leq$ $M_V\leq +3$, $-90^\circ \leq b \leq -75^\circ$, $V \leq 11$ and $1.0 \leq B-V \leq 1.5$. This gave $\sim$ 3000 stars. For the BSC giant simulation, we chose stars within a distance of 0.6 kpc, $-3 \leq$ $M_V \leq +3$, $V \leq 7$ and $0.8 \leq B-V \leq 1.8$. No cuts were made in position on the sky. This gave $\sim 10,000$ stars. 
Fig. 6 compares the luminosity functions of our sample of stars (FF and BSC: upper panel) with the luminosity function of the simulation. The red clump stars ($M_V$ $\sim$ 0.8) stand out very clearly in the Besan\c{c}on model
simulation as well as in our sample, as expected. All figures involving the Besan\c{c}on model in the main body of the paper are using this combined simulated catalog of the FF and BSC red giants.

\begin{figure} 
\includegraphics[width=0.495\textwidth]{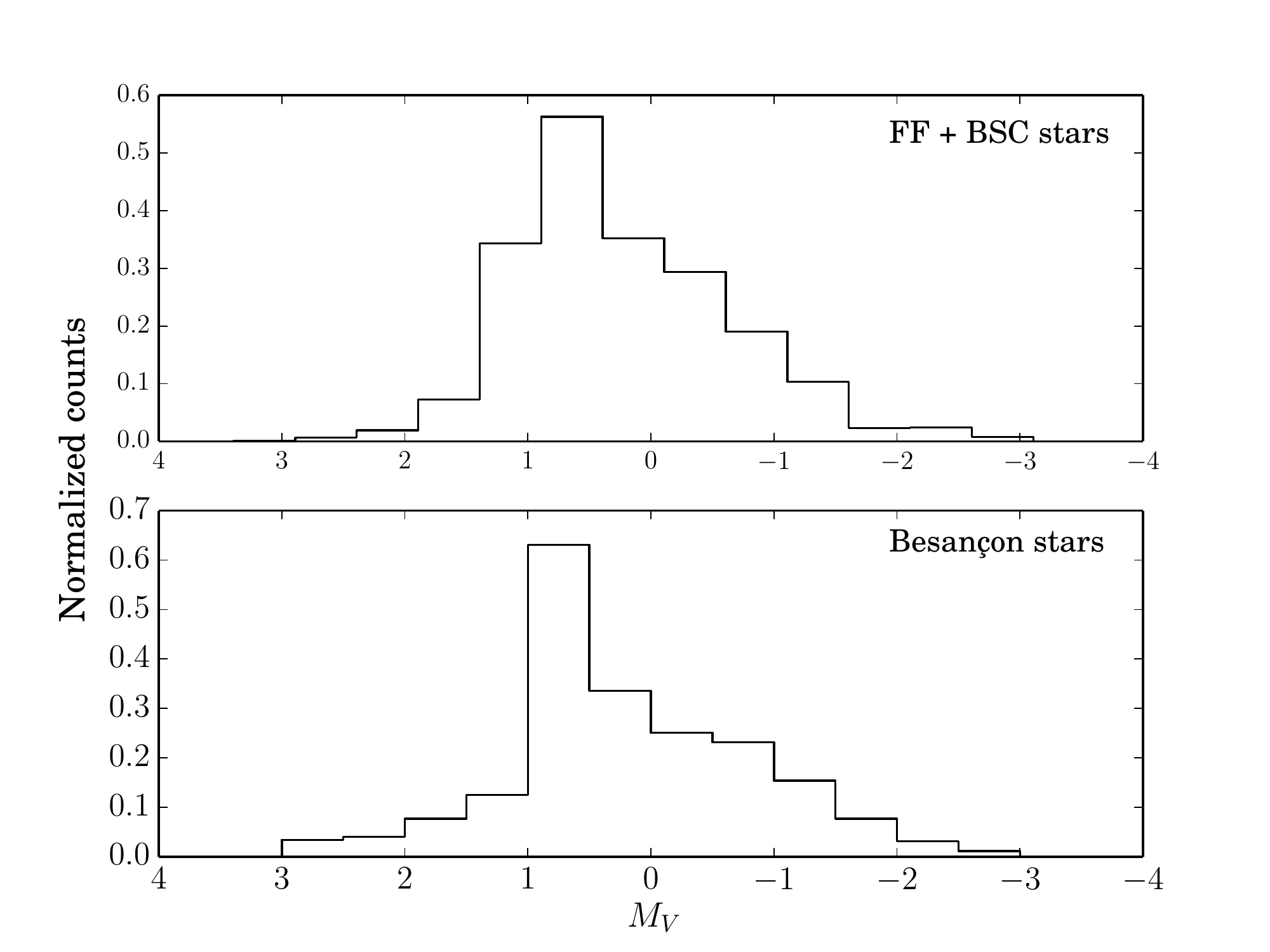}
\caption{Comparion of the luminosity functions of our sample of stars with a simulation of these stars from the Besan\c{c}on model. There seems to be a larger fraction of stars with $M_V \sim$ 1 in our sample, but otherwise the two samples are a good match. }
\end{figure}

Fig. 7 shows the W velocity vs absolute magnitude for the Besan\c{c}on sample of stars. As in Fig. 2, a kinematically colder component of bright giants can be clearly seen at $M_V < -2$. A small number of high velocity thick disc stars are present for $M_V > -2$.
 Fig. 8 is similar to Fig. 3 but for the Besan\c{c}on sample of stars.
\begin{figure} 
\includegraphics[width=0.495\textwidth]{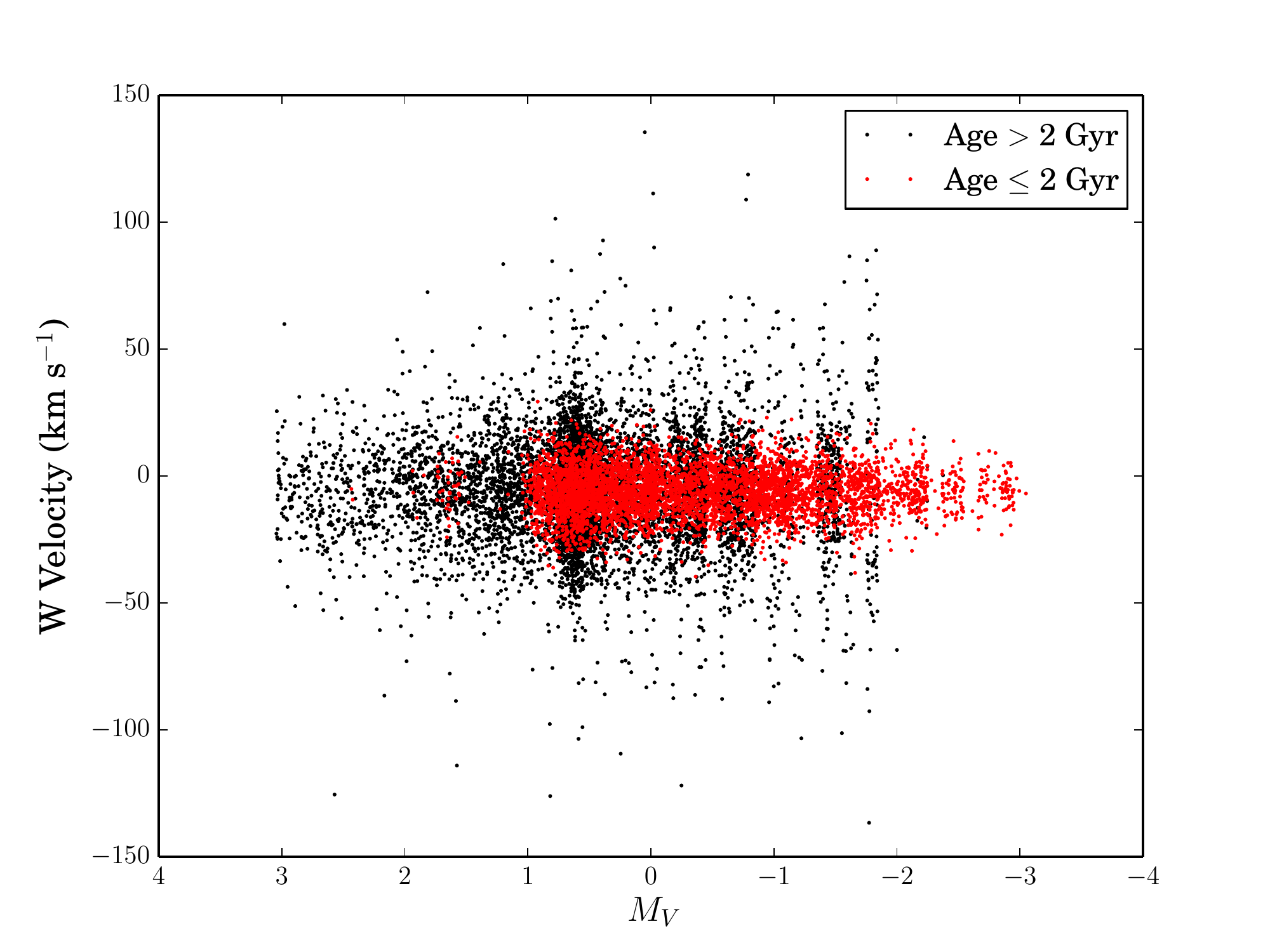}
\caption{Besan\c{c}on stars. Stars with $M_V \leq -2$ is comprised purely of the young population. A mixed population of young and old stars can be at $-2 < M_V \leq -1$.}
\end{figure}

\begin{figure} 
\includegraphics[width=0.495\textwidth]{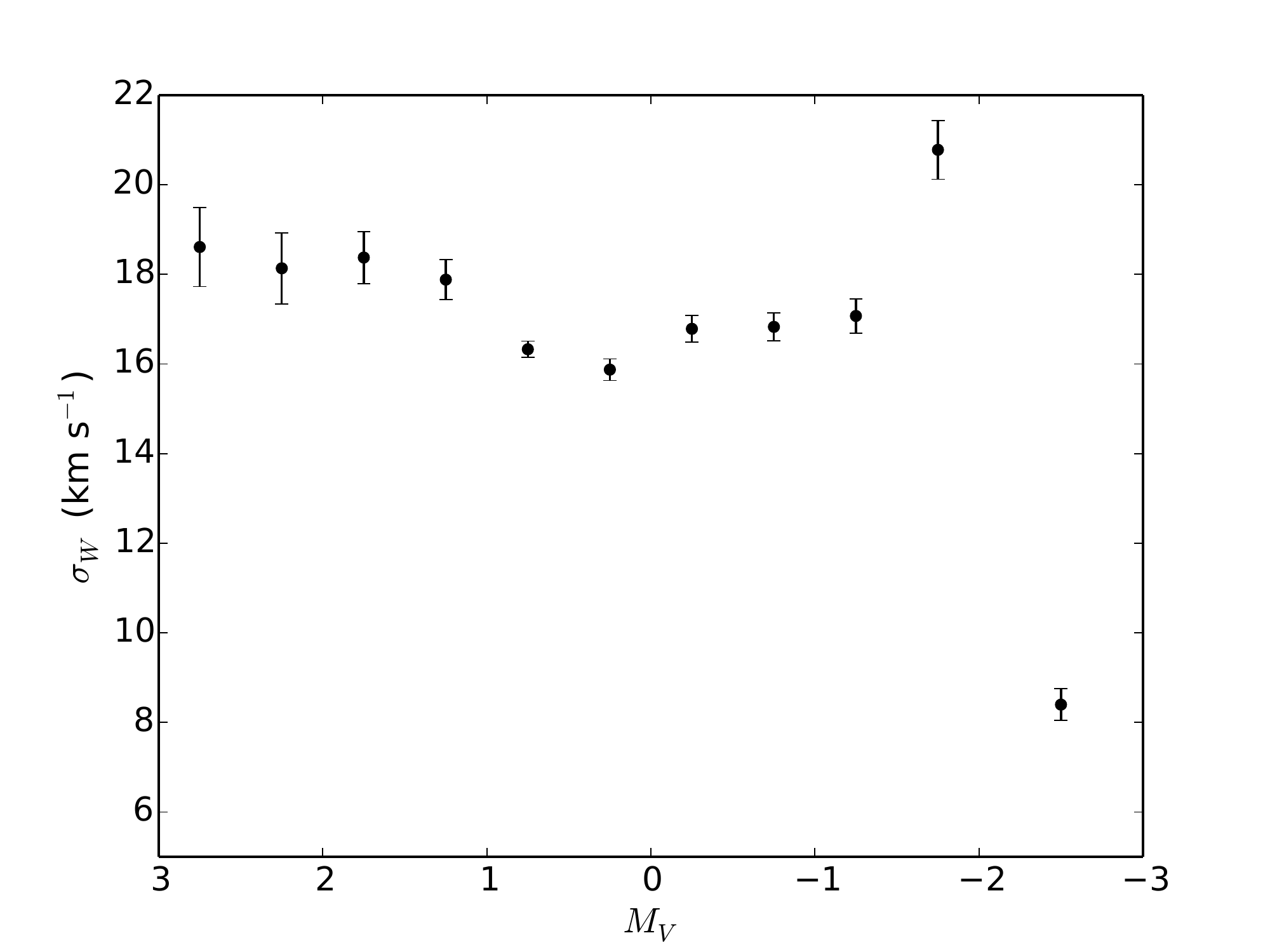}
\caption{The vertical velocity dispersion ($\sigma_W$) versus absolute magnitude for the Besan\c{c}on sample. The brighter stars ($M_V < -2$) are much colder as indicated by their low $\sigma_W$ value.}
\end{figure}

Fig. 9 is similar to Fig. 5 but for the Besan\c{c}on simulation. As before, the brightest stars correspond to the kinematically cold component. The slope of the linear fit corresponds to a velocity dispersion $\sigma$ = 7.9 \kms\ for the stars with $M_V \leq -1.8$. The two panels of stars with $M_V > -1.8$ contain stars of all ages and have again similar larger dispersions. 

\begin{figure} 
\includegraphics[width=0.49\textwidth]{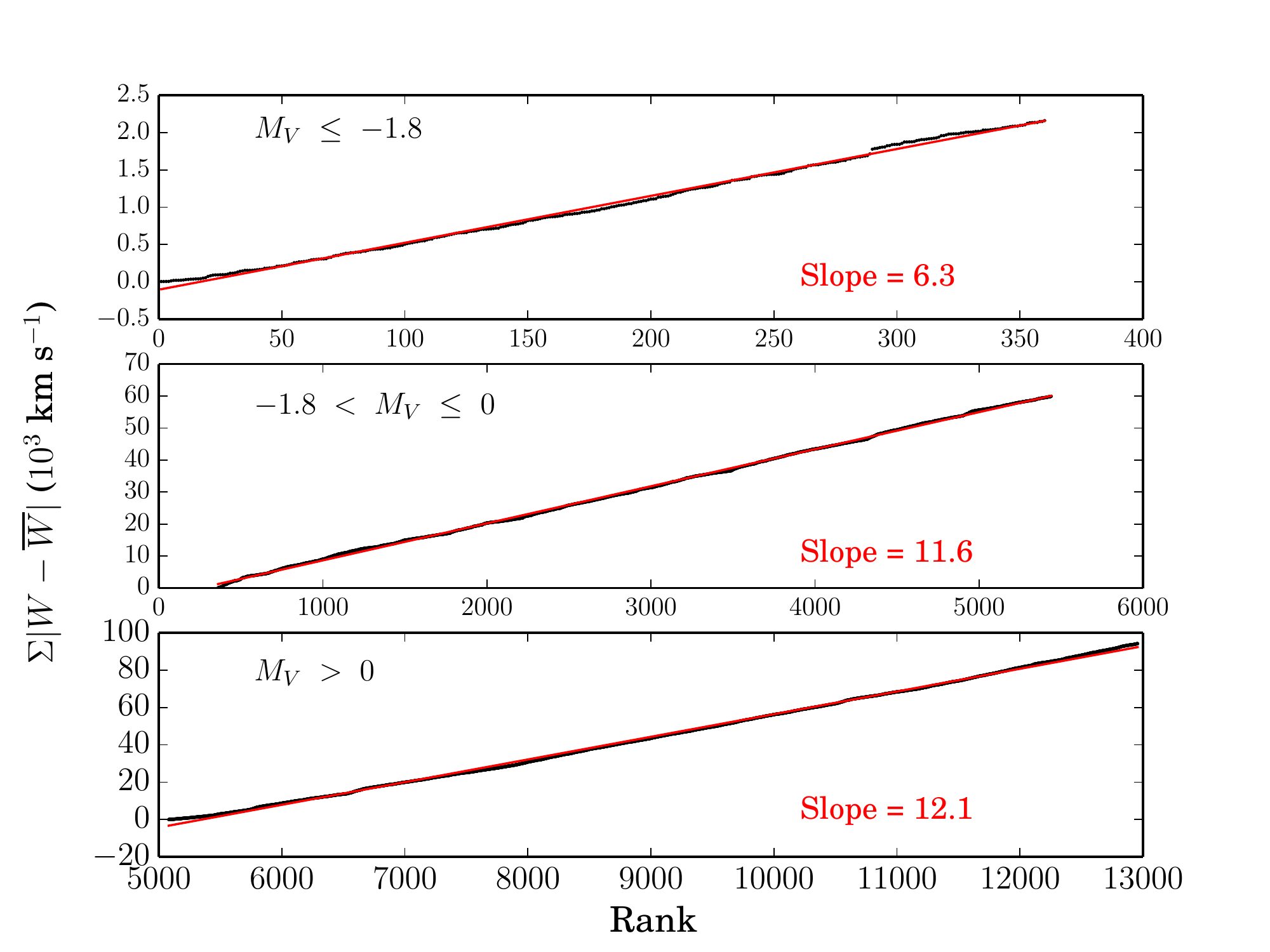}
\caption{The cumulative W velocities in different magnitude ranges of the Besan\c{c}on simulation. The x axis shows the rank of the stars and the y axis is the sum of modulus of $W - \overline{W}$ velocities. See text for further details. }
\end{figure}

We made a similar analysis to that shown in Fig. 4, using the W velocities from the Besan\c{c}on simulations to create a generalized histogram of velocities, and adopting a velocity error of $2$ \kms\ for each star. The results are tabulated in Table 3. 
\begin{table} 
\resizebox{0.48\textwidth}{!}{%
\begin{tabular}{cccc}
\hline
 & Cold Component & Hot Component & Single Gaussian Fit \\ \hline
Amplitude & $264.3 \pm 3.8$ & $148.0 \pm 4.0$ & $394.4 \pm 1.5$ \\ 
Mean (km s$^{-1}$) & $-5.64 \pm 0.03$ & $-6.26 \pm 0.08$ & $-5.82 \pm 0.05$ \\ 
Sigma (km s$^{-1}$) & $8.95 \pm 0.07$ & $18.45 \pm 0.17$ & $13.0 \pm 0.1$ \\ \hline
\end{tabular}
}
\caption{The fit results for the 1 component and 2 component models for the Besan\c{c}on sample of stars.}
\end{table}

Fig. 10 shows the velocity distribution for the Besan\c{c}on simulation fit with a double Gaussian model and with a single Gaussian model. Clearly, the double Gaussian is a better fit to the data. 

The Besan\c{c}on simulation is a good match to the kinematics of our sample of nearby giants. The two-component dispersion values are very similar to the values obtained for the FF + BSC stars. The relative numbers of stars in the two components can be calculated from the ratio of the product of the amplitude and dispersion $A\,\sigma$ for the two components (see Eqn. 2): the ratio of hot:cold stars is about 1.2:1 in both the FF + BSC sample and in the Besan\c{c}on simulation.

\begin{figure} 
\includegraphics[width=0.5\textwidth]{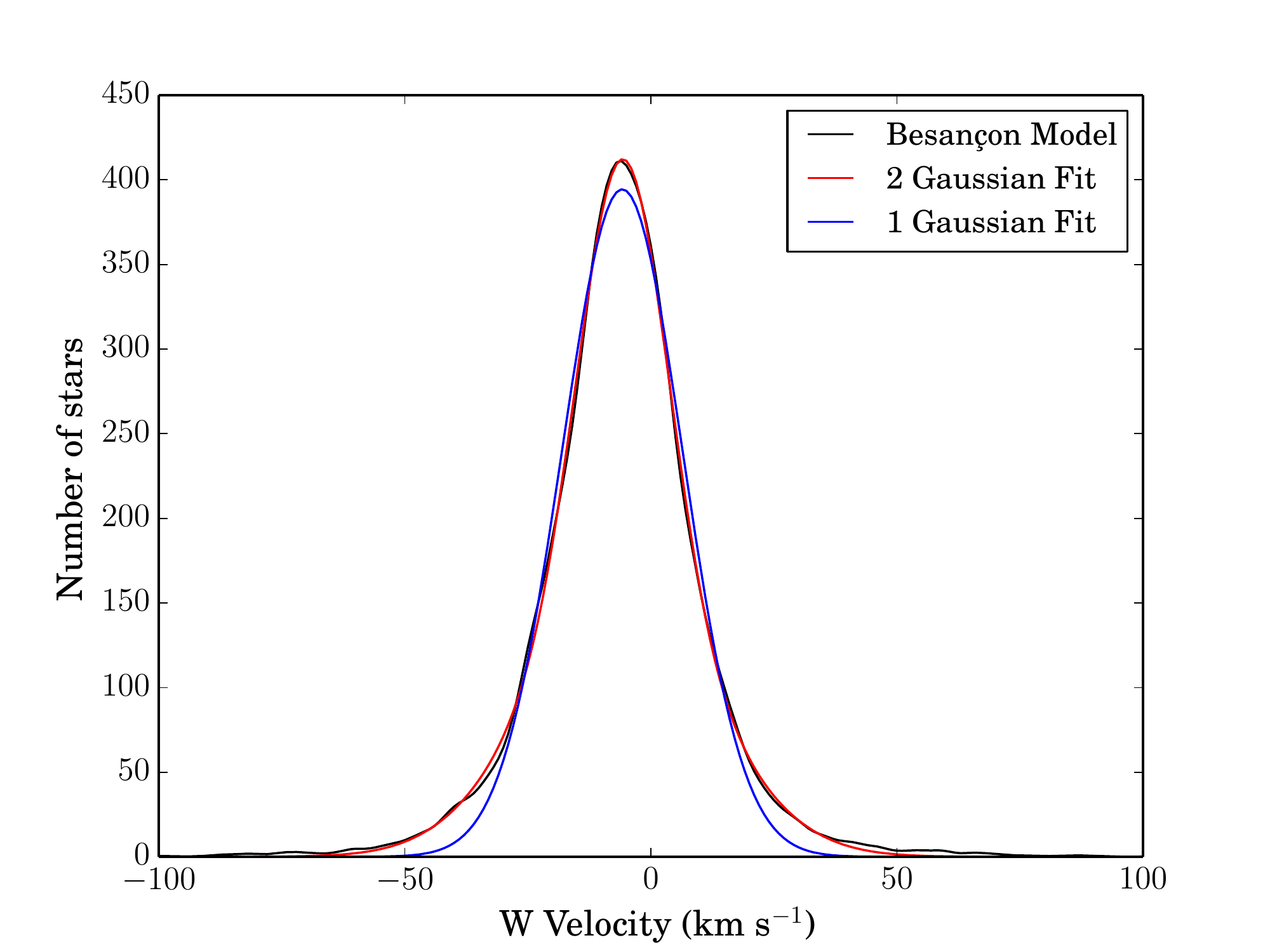}
\caption{Generalized histogram representing the stars simulated using the Besan\c{c}on model. The black curve shows the histogram and the red curve is the 2-component fit. }
\end{figure}

In evaluating the contribution of the kinematically cold component to the integrated light spectrum of a disc with a star formation history like that of the solar neighbourhood, it would be useful to know the ratio of the relative contributions of the colder and hotter components to the surface brightness of the disc. This would allow us to evaluate how much each component contributes to the integrated light of the disc. To estimate this ratio, we built up another velocity histogram using the Besan\c{c}on model, but this time each star was represented as a Gaussian with area proportional to its luminosity. This histogram, shown in Fig. 11, is then a luminosity-weighted
velocity distribution. The luminosity of each star was calculated from its $M_V$ value in the Besan\c{c}on simulation of the FF and BSC catalogs described earlier in this section. As before, we fit 2 Gaussians to the generalized histogram. The fit results are shown in Table 4.
\begin{table} 
\resizebox{0.48\textwidth}{!}{%
\begin{tabular}{cccc}
\hline
 & Cold Component & Hot Component & Single Gaussian Fit \\ \hline
Amplitude & $45700 \pm 400$ & $12000 \pm 500$ & $55500 \pm 300$ \\ 
Mean (km s$^{-1}$) & $-5.73 \pm 0.03$ & $-6.078 \pm 0.16$ & $-5.788 \pm 0.062$ \\ 
Sigma (km s$^{-1}$) & $8.23 \pm 0.05$ & $19.79 \pm 0.30$ & $10.265 \pm 0.062$ \\ \hline
\end{tabular}
}
\caption{Results of the Gaussian fit for the luminosity weighted generalized histogram of the Besan\c{c}on sample of stars.}
\end{table}

The total luminosity of each component is again proportional to the product $A\,\sigma$. Although the colder stars are fewer in number than the hotter stars, we find that their luminosity is a factor of $1.58 \pm 0.06$ higher than for the hotter stars. This younger and kinematically colder component would thus dominate the integrated light of a disc galaxy for which the star formation history of the thin disc was similar to that adopted in the Besan\c{c}on model. We also tried to use similar methods to obtain the luminosity ratio of the kinematically cold/hot stars from our sample (FF \& BSC) of stars but, with the luminosity weighting and the small sample size, we were not able to recover the older disc component of stars. The order of magnitude larger sample size of the Besan\c{c}on simulation allowed us to derive the luminosity ratio without difficulty. \\
\begin{figure}
\includegraphics[width=0.50\textwidth]{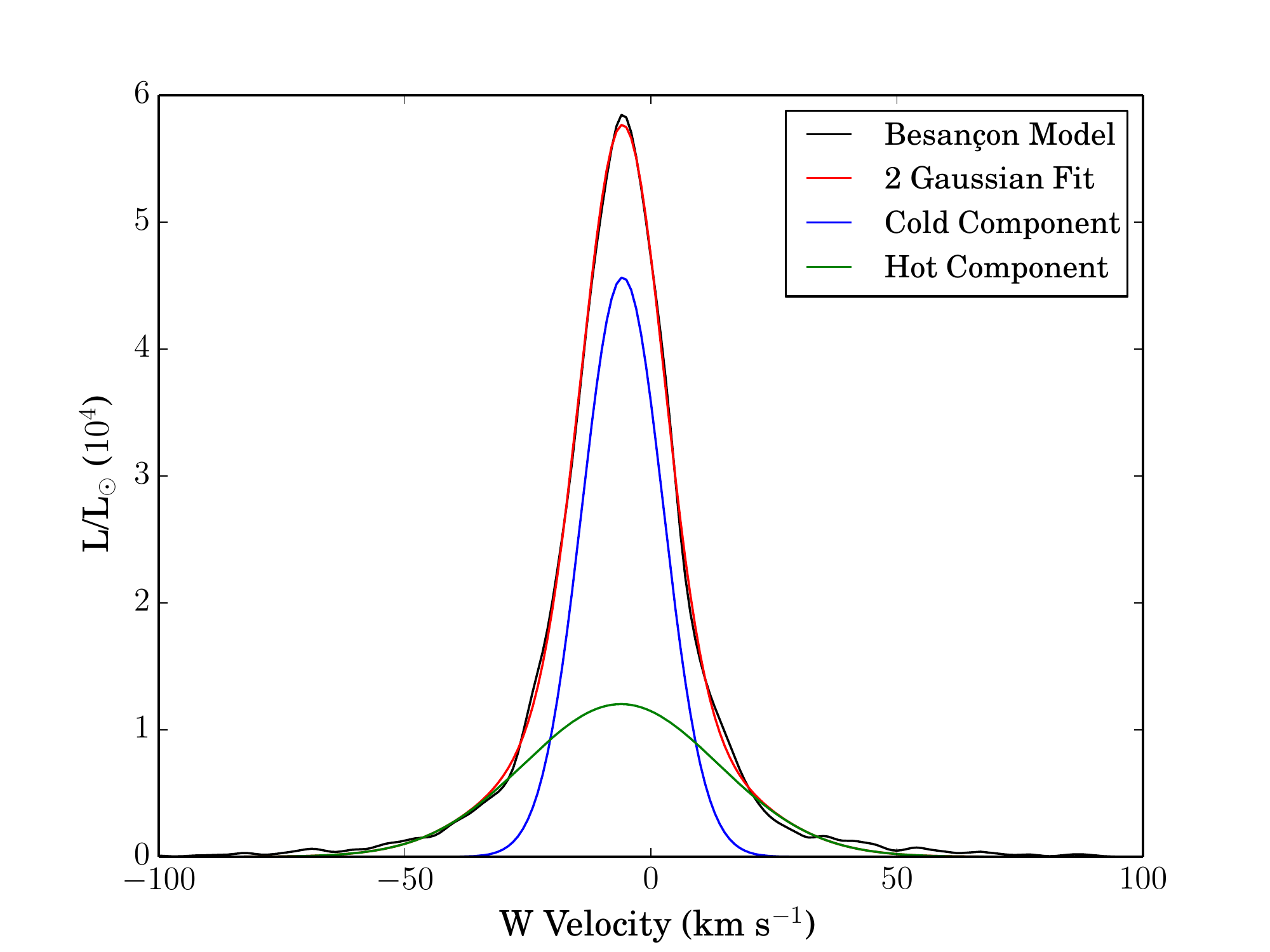}
\caption{Generalized histogram of the W-distribution of the stars from the Besan\c{c}on model. Each star contributed a Gaussian with area proportional to its luminosity and dispersion of $2$ \kms. The red curve is the two component fit to the Besan\c{c}on sample in black. The younger component and the older component are shown in blue and green respectively. }
\end{figure}

\begin{figure}
\includegraphics[width=0.49\textwidth]{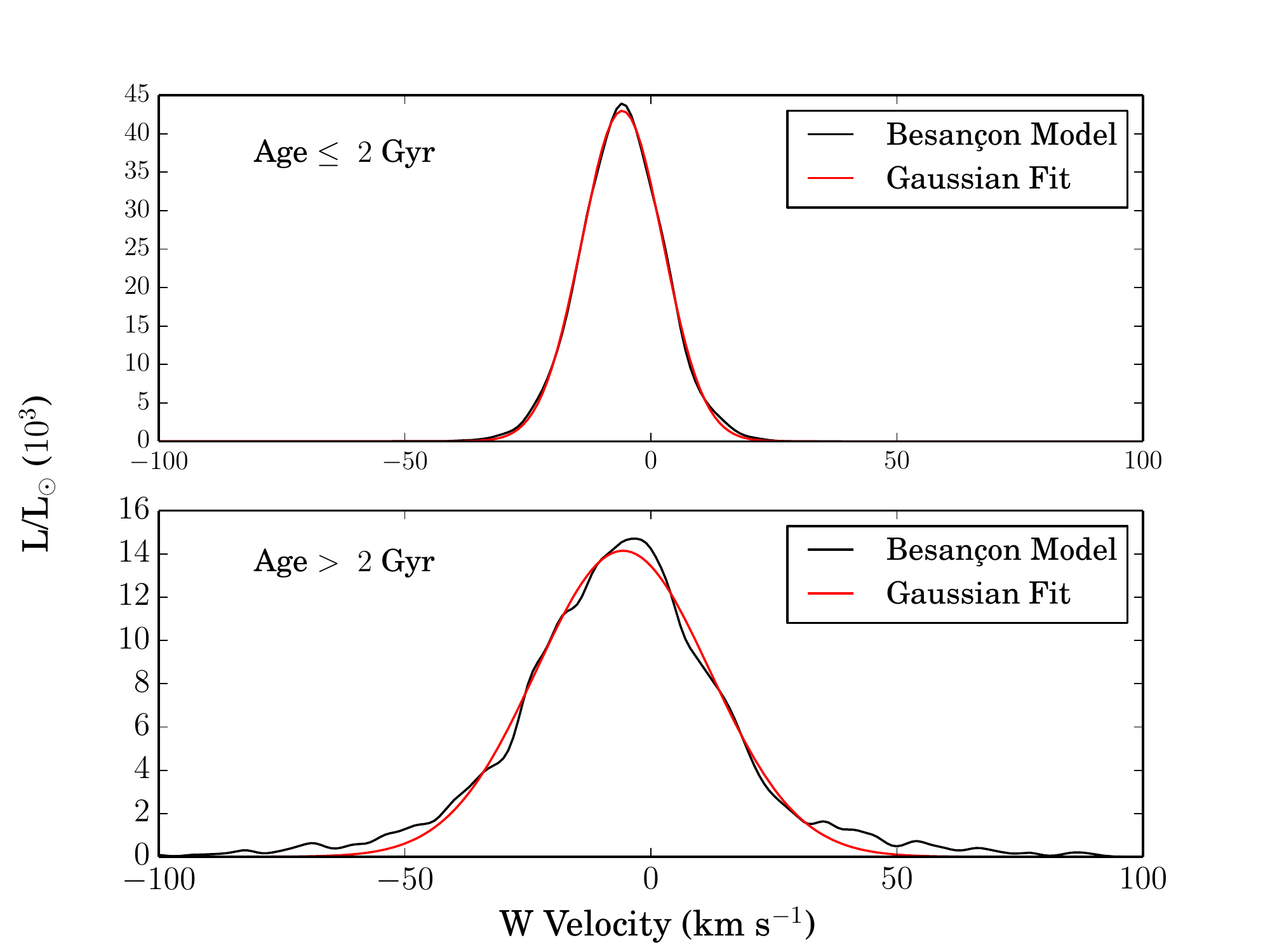}
\caption{Generalized luminosity histogram of the W-distribution of the stars from the Besan\c{c}on model based on age. The upper panel has all stars of age $\leq$ 2 Gyrs. These stars contribute the colder component with a narrow $\sigma_W$. The bottom panel shows the older stars that can be represented with a Gaussian with larger $\sigma \sim$ 18 km s$^{-1}$. }
\end{figure}

Fig. 12 shows the Besan\c{c}on stars split by age into 2 samples. 
The younge stars (age $\leq$ 2 Gyr) have a small dispersion of 8.238 $\pm$ 0.013 km s$^{-1}$ and the older stars have a larger dispersion of 17.695 $\pm$ 0.096 km s$^{-1}$, as expected.

\section{IMPLICATIONS FOR EXTERNAL GALAXIES}
The Besan\c{c}on simulations in the previous section looked at stars in a cone. In order to mimic observing the velocity distribution in external galaxies, we need a sample of stars in a cylinder perpendicular to the plane
of the disc. We used the Besan\c{c}on model to choose giants with spectral type G8III - K5III in the entire simulated sky, with $-3 \ \leq M_V \ \leq +3$ and $0.8 \ \leq B-V \ \leq \ 1.8$, within a distance of 5 kpc. We then selected giants in a cylinder of radius 2 kpc. This sample would be similar to what we would observe in the disc of a face-on galaxy from afar.

Fitting a double and single Gaussian to the luminosity-weighted velocity histogram of these sample of stars give us the results shown in Table 5. The fit is shown in Fig. 13.
\begin{table} 
\resizebox{0.48\textwidth}{!}{%
\begin{tabular}{cccc}
\hline
 & Cold Component & Hot Component & Single Gaussian Fit \\ \hline
Amplitude ($\times 10^3$) & $2050 \pm 26$ & $650 \pm 27$ & $2633 \pm 7$ \\ 
Mean (km s$^{-1}$) & $-5.96 \pm 0.03$ & $-5.81 \pm 0.11$ & $-5.93 \pm 0.05$ \\ 
Sigma (km s$^{-1}$) & $13.0 \pm 0.07$ & $23.90\pm 0.30$ & $15.40 \pm 0.05$ \\ \hline
\end{tabular}
}
\caption{Gaussian fit to the luminosity-weighted histogram of the Besan\c{c}on sample of stars in a cylinder.}
\end{table}
\begin{figure}
\includegraphics[width=0.49\textwidth]{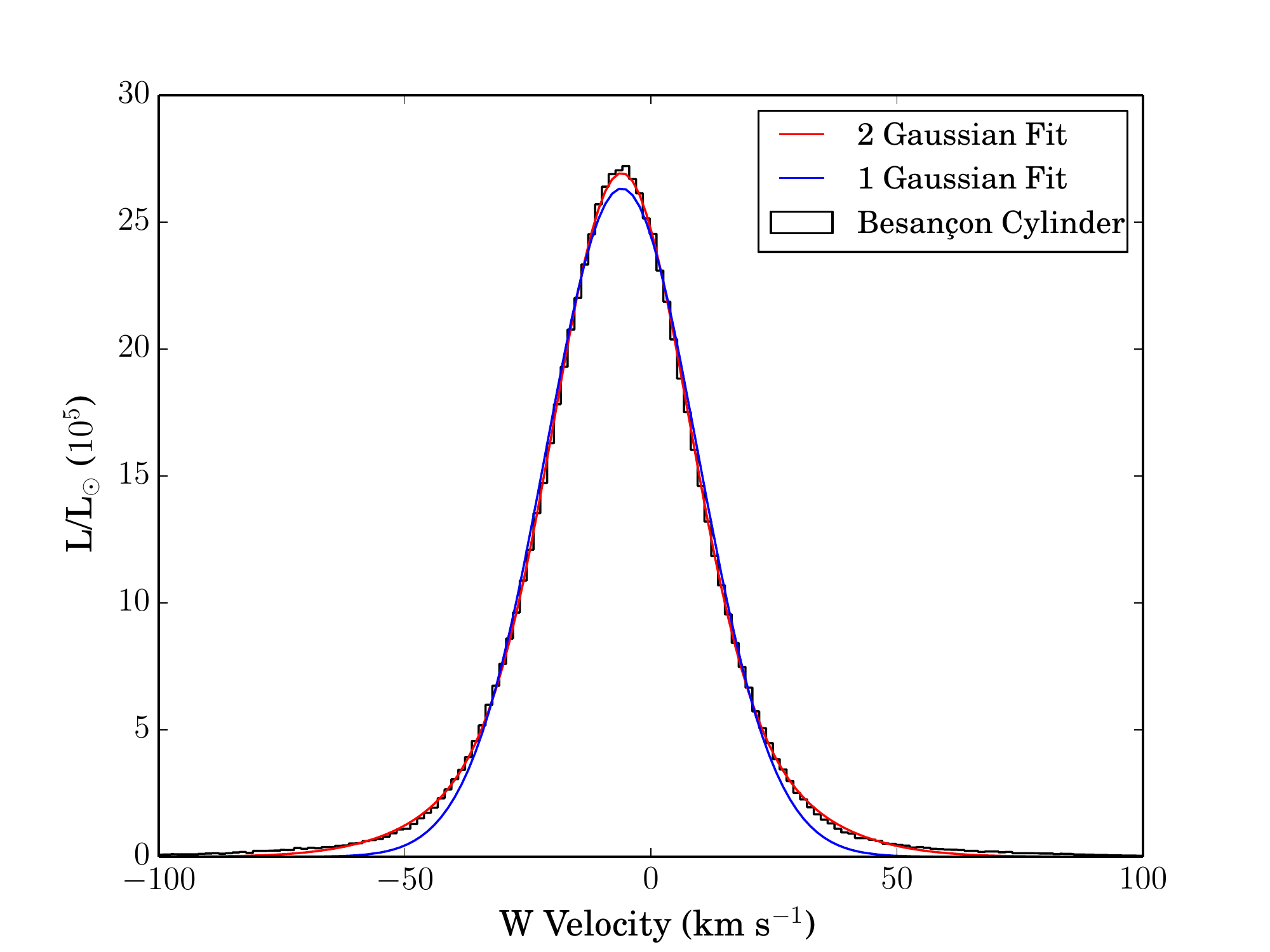}
\caption{Luminosity-weighted histogram of the W-distribution of the stars from the Besan\c{c}on stars in a cylinder. The double Gaussian fit is visibly better than the single Gaussian fit.}
\end{figure}
Using an analysis similar to that in Section 5, we find that the ratio of the estimated surface mass density of the disc between using the dispersion of the hotter thin disc component and using the dispersion for the single Gaussian fit, is:
\begin{equation}
\frac{\Sigma_{2}}{\Sigma_{1}} = \frac{\sigma_2^2}{\sigma_1^2} = \frac{(23.9 \pm 0.3)^2}{(15.40 \pm 0.05)^2} = 2.41 \pm 0.06
\end{equation}

\noindent As before, the subscript 1 refers to the single Gaussian fit and subscript 2 refers to the hotter (older) component of the double Gaussian fit. 
The underestimation of the disc surface mass density in the cylindrical geometry is similar to the value that we got for our nearby sample of stars (see section 4). 

The velocity dispersions that we calculated above are appropriate to a region with a similar star formation history and dynamical history to the solar neighbourhood. To get an idea of how the dispersions of the various components would change with a different star formation history, we constructed two different samples of stars: one with half the star formation rate over the last Gyr, and one with no star formation over the last Gyr. We again used a luminosity-weighted histogram to determine how the vertical velocity dispersions of the different Gaussian components would change. Our results are tabulated in Table 6. 
\begin{table} 
\resizebox{0.5\textwidth}{!}{%
\begin{tabular}{ccc}
\hline
&No star formation over last Gyr & Half the star formation over last Gyr \\ 
& $\sigma$ (km s$^{-1}$) & $\sigma$ (km s$^{-1}$)  \\\hline
Cold Component & $14.06 \pm 0.05$ & $13.56 \pm 0.06$ \\
Hot Component & $27.74 \pm 0.36$ & $25.70 \pm 0.33$ \\ 
Single Gaussian Fit & $15.79 \pm 0.05$ & $15.61 \pm 0.05$ \\ \hline
\end{tabular}
}
\caption{Results of the Gaussian fit for the luminosity-weighted histogram of the Besan\c{c}on sample of stars in a cylinder with different star formation histories.}
\end{table}
Although the hot and cold components become slightly hotter as the recent star formation rate decreases, the dispersions for the single Gaussian fits change very little with the change in star formation history. This
reflects the age velocity relation adopted in the Besan\c{c}on model.

\section{CONCLUSIONS}
Our analysis of W velocities of red giants in the solar neighbourhood shows the presence of two different population of stars -- a kinematically hotter component with a dispersion of $18.6 \pm 1.0$ \kms\ and a kinematically colder component with a dispersion of $9.6 \pm 0.5$ \kms. By number, the ratio of hot to cold stars in the solar neighbourhood is about 1.2:1, but this ratio would depend on the star formation history and dynamical evolution of the particular region that is under investigation. We note that the outcome of a simulated sample from the Besan\c{c}on model, using similar selection criteria, is an excellent match to our stars both in the velocity dispersions and the ratios of hot to cold stars.

If we were to regard the giants as a single kinematically homogeneous population, and if we further assumed that the scale height of this population is the scale height derived statistically for the old disc of the galaxy, then we would underestimate the surface mass density of the disc by a factor of $2.05 \pm 0.16$ (section 4). We get a similar value of $2.41 \pm 0.06$ when looking at stars in a cylinder in the Besan\c{c}on model (section 6). These assumptions are usually made in making kinematical estimates of the surface density of external disc galaxies. 
Again, this underestimate of a factor of 2 is appropriate for a region of a disc galaxy which has a similar star formation history and dynamical history to the Galactic disc near the sun. Using a velocity  dispersion of $18.6$ km s$^{-1}$ and a scale height from literature, we obtain a disc surface density $\Sigma = 57 \pm 12$ M$_\odot$ pc$^{-2}$, which is in fair agreement with the range of estimates calculated for the Milky Way in previous studies, using alternative techniques.

This work demonstrates an issue for studies that use the velocity dispersion of stars in the disc of external near-face-on galaxies to estimate their surface mass density. The scale height for these galaxies come from red or near-infrared photometry that is sensitive to the kinematically hotter population of disc stars. On the other hand, the integrated light spectroscopy that is used to estimate the vertical velocity dispersions includes contributions from both younger and older populations of star. We have shown that the contribution of the kinematically colder stars to the integrated light of the solar neighbourhood may significantly outweigh the contribution from the kinematically hotter old disc, by a factor of about 1.56 in surface brightness. This factor is about 1.7 for the
cylindrical sample from the Besan\c{c}on Model (Section 6). The outcome would be a significant underestimate of the velocity dispersion of the old disc, and hence of the surface density of the disc. Thus a maximal disc would appear submaximal, and its dark halo will appear to have a shorter scale length and higher central density than its true value.

We can illustrate this effect quantitatively with a decomposition of the rotation curve of the spiral galaxy NGC 3198 which was studied in detail by \citet{vanAlbada}. The upper panel of Fig. 14 shows a maximum disc decomposition similar to that shown in Fig. 4 of \citet{vanAlbada}. We have used the disc model of \citet{vanAlbada} and the widely used pseudo-isothermal-sphere (PITS) dark halo model with $$\rho(r)/\rho(0) = [1 + (r/a)^2]^{-1}$$ for consistency with later work such as \citet{Kormendy14}.  The rotation curve for the PITS dark halo model has the form $$V_c^2(r) = V_\infty^2 [1 - (\tan^{-1}x)/x],$$ where
$V_\infty^2 = 4 \pi G \rho(0) a^2$ and $x = r/a$.  The adopted dark halo model parameters for the maximum disc decomposition are: $a = 7.45$ kpc, $V_\infty = 161$ km s$^{-1}$ and a central density $\rho(0) = 0.0086\, M_\odot$ pc$^{-3}$. Although our dark halo model is slightly different from the model used by \citet{vanAlbada} (both have constant
density central cores), the halo $V_c$ curves are almost identical.  The lower panel of Fig. 14 shows the decomposition for a disc surface density that is lower by a factor $2$ than the one shown in the upper panel. This decomposition and the maximum disc decomposition are equally satisfactory, illustrating the disc-halo degeneracy, but the amplitude of the rotation contribution for the disc is now lower by $\sqrt 2$ and the required halo has a much shorter scale length $a$ and higher central density $\rho(0)$. The halo model shown in the lower panel of Fig. 14 has $a = 1.5$ kpc, $V_\infty = 144$ km s$^{-1}$ and $\rho(0) = 0.17\, M_\odot$ pc$^{-3}$. In summary, if a disc of a galaxy like NGC 3198 is truly maximal and its surface density is underestimated by a factor of 2, as we believe could follow from adopting a single velocity dispersion for the young and old giants, then the central density of its dark halo would be overestimated by a factor of about $20$ and its scale length $a$ would be underestimated by a factor of about $5$.  This would introduce a serious distortion into the scaling laws for dark haloes (see e.g. \citealt{Kormendy14} ).

\begin{figure}
\includegraphics[width=0.49\textwidth]{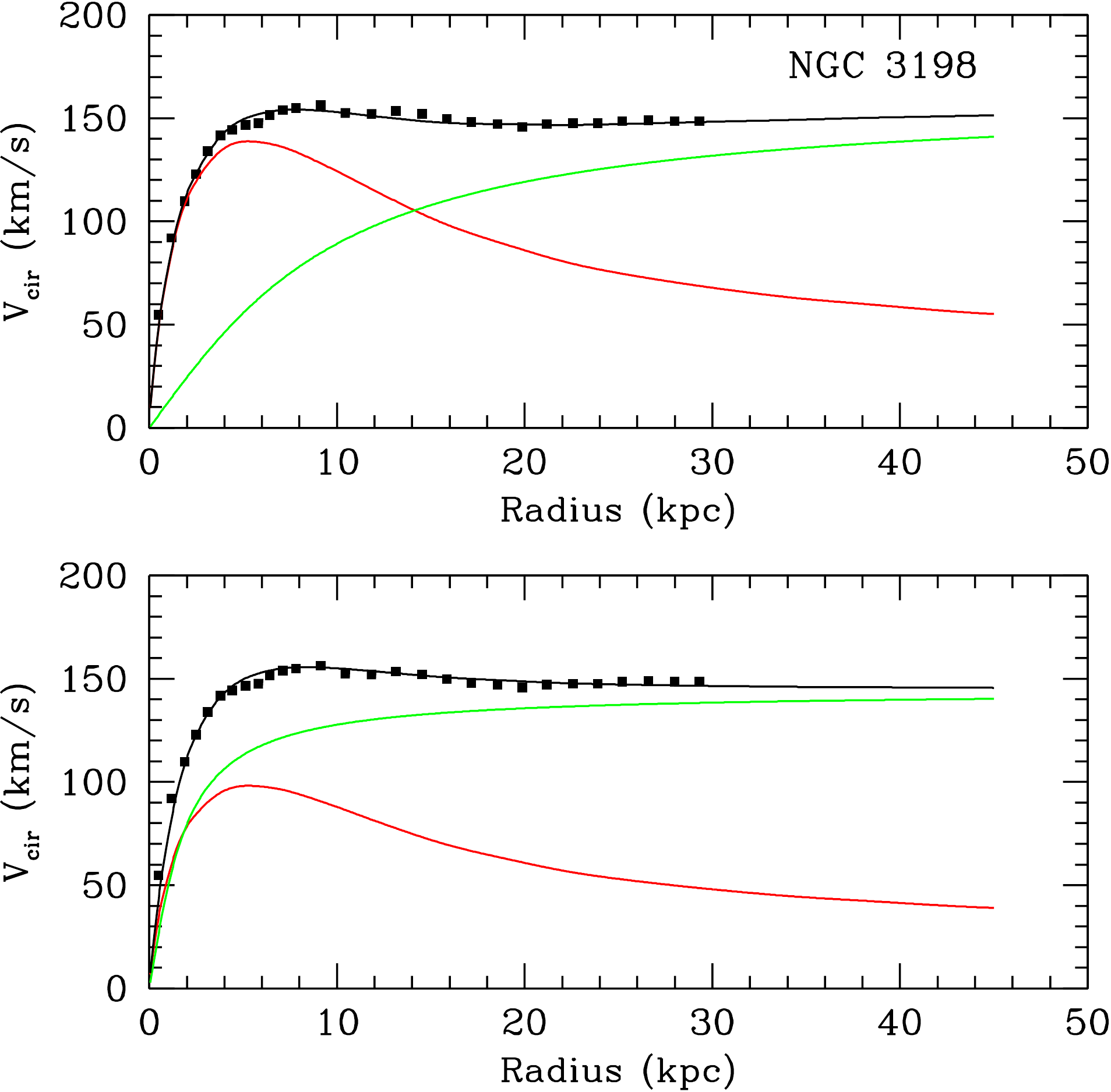} 
\caption{Upper panel shows a maximum disc decomposition of the rotation curve of NGC 3198. Rotation data (black squares) and the rotation curve for the maximum disc (red curve) are adapted from \citet{vanAlbada}. The green curve shows the rotation curve for the required halo, and the black curve is the total rotation curve from disc + halo (see section 7 for details). Lower panel shows the decomposition for a disc with surface density $= 0.5$ of the maximum disc value. The required dark halo is now much more compact: its central density is about $20$ times higher than for the maximum disc decomposition and its scale length is about $5$ times shorter.}
\end{figure}

\section{Future Work}

In order to handle the problem of kinematical inhomogeneity described here, we are undertaking a program to measure the velocity dispersion of the old disc of several large nearby disc galaxies, by making (i) a two-component analysis of the integrated light spectra of regions in their discs, and (ii) measuring radial velocities for large numbers of planetary nebulae in their discs. As for the red giants, the planetary nebulae have progenitors of a wide range of ages, and we can expect them to include kinematically hot and cold sub-components. The primary outcome will be improved surface densities of the stellar discs, which will then be used to constrain the decomposition of the HI rotation curves into contributions from the disc and the dark halo.

\section{Acknowledgements}
The authors would like to thank the referee for useful comments that helped improve this paper.
The authors are grateful to Marc Verheijen and Matthew Bershady for helpful discussions. SA would like to thank ESO for the ESO studentship that supported part of this work. KCF thanks Piet van der Kruit for many discussions on related topics over many years. This work has made use of the 
Besan\c{c}on Model and the IAC-STAR Synthetic CMD computation code. IAC-STAR is suported and maintained by the computer division of the Instituto de Astrofisica de Canarias.

\bibliographystyle{mnras}
\bibstyle{mnras}
\bibliography{Surya}

\begin{thebibliography}{}
\makeatletter
\relax
\def\mn@urlcharsother{\let\do\@makeother \do\$\do\&\do\#\do\^\do\_\do\%\do\~}
\def\mn@doi{\begingroup\mn@urlcharsother \@ifnextchar [ {\mn@doi@}
  {\mn@doi@[]}}
\def\mn@doi@[#1]#2{\def\@tempa{#1}\ifx\@tempa\@empty \href
  {http://dx.doi.org/#2} {doi:#2}\else \href {http://dx.doi.org/#2} {#1}\fi
  \endgroup}
\def\mn@eprint#1#2{\mn@eprint@#1:#2::\@nil}
\def\mn@eprint@arXiv#1{\href {http://arxiv.org/abs/#1} {{\tt arXiv:#1}}}
\def\mn@eprint@dblp#1{\href {http://dblp.uni-trier.de/rec/bibtex/#1.xml}
  {dblp:#1}}
\def\mn@eprint@#1:#2:#3:#4\@nil{\def\@tempa {#1}\def\@tempb {#2}\def\@tempc
  {#3}\ifx \@tempc \@empty \let \@tempc \@tempb \let \@tempb \@tempa \fi \ifx
  \@tempb \@empty \def\@tempb {arXiv}\fi \@ifundefined
  {mn@eprint@\@tempb}{\@tempb:\@tempc}{\expandafter \expandafter \csname
  mn@eprint@\@tempb\endcsname \expandafter{\@tempc}}}

\bibitem[\protect\citeauthoryear{{Anderson} \& {Francis}}{{Anderson} \&
  {Francis}}{2012}]{XHIP}
{Anderson} E.,  {Francis} C.,  2012, \mn@doi [Astronomy Letters]
  {10.1134/S1063773712050015}, \href
  {http://adsabs.harvard.edu/abs/2012AstL...38..331A} {38, 331}

\bibitem[\protect\citeauthoryear{{Aparicio} \& {Gallart}}{{Aparicio} \&
  {Gallart}}{2004}]{IAC}
{Aparicio} A.,  {Gallart} C.,  2004, \mn@doi [AJ] {10.1086/382836}, \href
  {http://adsabs.harvard.edu/abs/2004AJ....128.1465A} {128, 1465}

\bibitem[\protect\citeauthoryear{{Bell} \& {de Jong}}{{Bell} \& {de
  Jong}}{2001}]{Bell:2001}
{Bell} E.~F.,  {de Jong} R.~S.,  2001, \mn@doi [ApJ] {10.1086/319728}, \href
  {http://adsabs.harvard.edu/abs/2001ApJ...550..212B} {550, 212}

\bibitem[\protect\citeauthoryear{{Bershady}, {Verheijen}, {Swaters},
  {Andersen}, {Westfall}  \& {Martinsson}}{{Bershady} et~al.}{2010}]{DMI}
{Bershady} M.~A.,  {Verheijen} M.~A.~W.,  {Swaters} R.~A.,  {Andersen} D.~R.,
  {Westfall} K.~B.,   {Martinsson} T.,  2010, \mn@doi [ApJ]
  {10.1088/0004-637X/716/1/198}, \href
  {http://adsabs.harvard.edu/abs/2010ApJ...716..198B} {716, 198}

\bibitem[\protect\citeauthoryear{{Bershady}, {Martinsson}, {Verheijen},
  {Westfall}, {Andersen}  \& {Swaters}}{{Bershady} et~al.}{2011}]{DM}
{Bershady} M.~A.,  {Martinsson} T.~P.~K.,  {Verheijen} M.~A.~W.,  {Westfall}
  K.~B.,  {Andersen} D.~R.,   {Swaters} R.~A.,  2011, \mn@doi [ApJ]
  {10.1088/2041-8205/739/2/L47}, \href
  {http://adsabs.harvard.edu/abs/2011ApJ...739L..47B} {739, L47}

\bibitem[\protect\citeauthoryear{{Bottema}}{{Bottema}}{1997}]{Bottema97}
{Bottema} R.,  1997, A\&A, \href
  {http://adsabs.harvard.edu/abs/1997A%26A...328..517B} {328, 517}

\bibitem[\protect\citeauthoryear{{Bottema}, {van der Kruit}  \&
  {Freeman}}{{Bottema} et~al.}{1987}]{Bottema}
{Bottema} R.,  {van der Kruit} P.~C.,   {Freeman} K.~C.,  1987, A\&A, \href
  {http://adsabs.harvard.edu/abs/1987A%26A...178...77B} {178, 77}

\bibitem[\protect\citeauthoryear{{Bovy} \& {Rix}}{{Bovy} \& {Rix}}{2013}]{BR}
{Bovy} J.,  {Rix} H.-W.,  2013, \mn@doi [ApJ] {10.1088/0004-637X/779/2/115},
  \href {http://adsabs.harvard.edu/abs/2013ApJ...779..115B} {779, 115}

\bibitem[\protect\citeauthoryear{{Casagrande}, {Sch{\"o}nrich}, {Asplund},
  {Cassisi}, {Ram{\'{\i}}rez}, {Mel{\'e}ndez}, {Bensby}  \&
  {Feltzing}}{{Casagrande} et~al.}{2011}]{Casagrande11}
{Casagrande} L.,  {Sch{\"o}nrich} R.,  {Asplund} M.,  {Cassisi} S.,
  {Ram{\'{\i}}rez} I.,  {Mel{\'e}ndez} J.,  {Bensby} T.,   {Feltzing} S.,
  2011, \mn@doi [A\&A] {10.1051/0004-6361/201016276}, \href
  {http://adsabs.harvard.edu/abs/2011A%26A...530A.138C} {530, A138}

\bibitem[\protect\citeauthoryear{{Delhaye}}{{Delhaye}}{1965}]{Delhaye65}
{Delhaye} J.,  1965, Stars and stellar systems.
 Vol. 6, Univ. of Chicago Press

\bibitem[\protect\citeauthoryear{{Edvardsson}, {Andersen}, {Gustafsson},
  {Lambert}, {Nissen}  \& {Tomkin}}{{Edvardsson} et~al.}{1993}]{Edvardsson93}
{Edvardsson} B.,  {Andersen} J.,  {Gustafsson} B.,  {Lambert} D.~L.,  {Nissen}
  P.~E.,   {Tomkin} J.,  1993, A\&A, \href
  {http://adsabs.harvard.edu/abs/1993A%26A...275..101E} {275, 101}

\bibitem[\protect\citeauthoryear{{Eriksson}}{{Eriksson}}{1995}]{Eriksson95}
{Eriksson} W.,  1995, {VizieR Online Data Catalog: Phot and Spectrophot
  Investigation, South Gal Pole (Eriksson 1978)}

\bibitem[\protect\citeauthoryear{{Flynn} \& {Freeman}}{{Flynn} \&
  {Freeman}}{1993}]{ff}
{Flynn} C.,  {Freeman} K.~C.,  1993, A\&AS, \href
  {http://adsabs.harvard.edu/abs/1993A%26AS...97..835F} {97, 835}

\bibitem[\protect\citeauthoryear{{Flynn} \& {Fuchs}}{{Flynn} \&
  {Fuchs}}{1994}]{Flynn:94}
{Flynn} C.,  {Fuchs} B.,  1994, MNRAS, \href
  {http://adsabs.harvard.edu/abs/1994MNRAS.270..471F} {270, 471}

\bibitem[\protect\citeauthoryear{{Flynn}, {Holmberg}, {Portinari}, {Fuchs}  \&
  {Jahrei{\ss}}}{{Flynn} et~al.}{2006}]{Flynn:06}
{Flynn} C.,  {Holmberg} J.,  {Portinari} L.,  {Fuchs} B.,   {Jahrei{\ss}} H.,
  2006, \mn@doi [MNRAS] {10.1111/j.1365-2966.2006.10911.x}, \href
  {http://adsabs.harvard.edu/abs/2006MNRAS.372.1149F} {372, 1149}

\bibitem[\protect\citeauthoryear{{Freeman}}{{Freeman}}{1991}]{Freeman91}
{Freeman} K.~C.,  1991, in {Sundelius} B.,  ed., Dynamics of Disc Galaxies.
  p.~15

\bibitem[\protect\citeauthoryear{{Gilmore} \& {Reid}}{{Gilmore} \&
  {Reid}}{1983}]{Gilmore:83}
{Gilmore} G.,  {Reid} N.,  1983, MNRAS, \href
  {http://adsabs.harvard.edu/abs/1983MNRAS.202.1025G} {202, 1025}

\bibitem[\protect\citeauthoryear{{Gomez}, {Grenier}, {Udry}, {Haywood},
  {Meillon}, {Sabas}, {Sellier}  \& {Morin}}{{Gomez} et~al.}{1997}]{Gomez97}
{Gomez} A.~E.,  {Grenier} S.,  {Udry} S.,  {Haywood} M.,  {Meillon} L.,
  {Sabas} V.,  {Sellier} A.,   {Morin} D.,  1997, in {Bonnet} R.~M.,  et~al.,
  eds,  ESA Special Publication Vol. 402, Hipparcos - Venice '97. pp 621--624

\bibitem[\protect\citeauthoryear{{Haywood}}{{Haywood}}{2008}]{Haywood08}
{Haywood} M.,  2008, \mn@doi [MNRAS] {10.1111/j.1365-2966.2008.13395.x}, \href
  {http://adsabs.harvard.edu/abs/2008MNRAS.388.1175H} {388, 1175}

\bibitem[\protect\citeauthoryear{{Herrmann}, {Ciardullo}, {Feldmeier}  \&
  {Vinciguerra}}{{Herrmann} et~al.}{2008}]{PNI}
{Herrmann} K.~A.,  {Ciardullo} R.,  {Feldmeier} J.~J.,   {Vinciguerra} M.,
  2008, \mn@doi [ApJ] {10.1086/589920}, \href
  {http://adsabs.harvard.edu/abs/2008ApJ...683..630H} {683, 630}

\bibitem[\protect\citeauthoryear{{Hoffleit} \& {Jaschek}}{{Hoffleit} \&
  {Jaschek}}{1991}]{BSC}
{Hoffleit} D.,  {Jaschek} C.~.,  1991, {The Bright star catalogue}

\bibitem[\protect\citeauthoryear{{Holmberg} \& {Flynn}}{{Holmberg} \&
  {Flynn}}{2004}]{Holmberg2004}
{Holmberg} J.,  {Flynn} C.,  2004, \mn@doi [MNRAS]
  {10.1111/j.1365-2966.2004.07931.x}, \href
  {http://adsabs.harvard.edu/abs/2004MNRAS.352..440H} {352, 440}

\bibitem[\protect\citeauthoryear{{Houk}}{{Houk}}{1982}]{houk}
{Houk} N.,  1982, {Michigan Catalogue of Two-dimensional Spectral Types for the
  HD stars. Volume\_3. Declinations $-40^\circ.0$ to $-26^\circ.0$.}

\bibitem[\protect\citeauthoryear{{Houk} \& {Smith-Moore}}{{Houk} \&
  {Smith-Moore}}{1988}]{houk2}
{Houk} N.,  {Smith-Moore} M.,  1988, {Michigan Catalogue of Two-dimensional
  Spectral Types for the HD Stars. Volume 4, Declinations $-26^\circ.0$ to
  $-12^\circ.0$.}

\bibitem[\protect\citeauthoryear{{Johnson} \& {Soderblom}}{{Johnson} \&
  {Soderblom}}{1987}]{JS}
{Johnson} D.~R.~H.,  {Soderblom} D.~R.,  1987, \mn@doi [AJ] {10.1086/114370},
  \href {http://adsabs.harvard.edu/abs/1987AJ.....93..864J} {93, 864}

\bibitem[\protect\citeauthoryear{{Just}, {Fuchs}, {Jahrei{\ss}}, {Flynn},
  {Dettbarn}  \& {Rybizki}}{{Just} et~al.}{2015}]{Just:15}
{Just} A.,  {Fuchs} B.,  {Jahrei{\ss}} H.,  {Flynn} C.,  {Dettbarn} C.,
  {Rybizki} J.,  2015, \mn@doi [MNRAS] {10.1093/mnras/stv858}, \href
  {http://adsabs.harvard.edu/abs/2015MNRAS.451..149J} {451, 149}

\bibitem[\protect\citeauthoryear{{Kormendy} \& {Freeman}}{{Kormendy} \&
  {Freeman}}{2014}]{Kormendy14}
{Kormendy} J.,  {Freeman} K.~C.,  2014, preprint, \href
  {http://adsabs.harvard.edu/abs/2014arXiv1411.2170K} {} (\mn@eprint {arXiv}
  {1411.2170})

\bibitem[\protect\citeauthoryear{{Kregel}, {van der Kruit}  \&
  {Freeman}}{{Kregel} et~al.}{2005}]{Kregel}
{Kregel} M.,  {van der Kruit} P.~C.,   {Freeman} K.~C.,  2005, \mn@doi [MNRAS]
  {10.1111/j.1365-2966.2005.08855.x}, \href
  {http://adsabs.harvard.edu/abs/2005MNRAS.358..503K} {358, 503}

\bibitem[\protect\citeauthoryear{{Kuijken} \& {Gilmore}}{{Kuijken} \&
  {Gilmore}}{1989a}]{KGI}
{Kuijken} K.,  {Gilmore} G.,  1989a, MNRAS, \href
  {http://adsabs.harvard.edu/abs/1989MNRAS.239..571K} {239, 571}

\bibitem[\protect\citeauthoryear{{Kuijken} \& {Gilmore}}{{Kuijken} \&
  {Gilmore}}{1989b}]{KGII}
{Kuijken} K.,  {Gilmore} G.,  1989b, MNRAS, \href
  {http://adsabs.harvard.edu/abs/1989MNRAS.239..605K} {239, 605}

\bibitem[\protect\citeauthoryear{{Kuijken} \& {Gilmore}}{{Kuijken} \&
  {Gilmore}}{1989c}]{KGIII}
{Kuijken} K.,  {Gilmore} G.,  1989c, MNRAS, \href
  {http://adsabs.harvard.edu/abs/1989MNRAS.239..651K} {239, 651}

\bibitem[\protect\citeauthoryear{{Macci{\`o}}, {Ruchayskiy}, {Boyarsky}  \&
  {Mu{\~n}oz-Cuartas}}{{Macci{\`o}} et~al.}{2013}]{Maccio:13}
{Macci{\`o}} A.~V.,  {Ruchayskiy} O.,  {Boyarsky} A.,   {Mu{\~n}oz-Cuartas}
  J.~C.,  2013, \mn@doi [MNRAS] {10.1093/mnras/sts078}, \href
  {http://adsabs.harvard.edu/abs/2013MNRAS.428..882M} {428, 882}

\bibitem[\protect\citeauthoryear{{Maraston}}{{Maraston}}{2005}]{Maraston05}
{Maraston} C.,  2005, \mn@doi [MNRAS] {10.1111/j.1365-2966.2005.09270.x}, \href
  {http://adsabs.harvard.edu/abs/2005MNRAS.362..799M} {362, 799}

\bibitem[\protect\citeauthoryear{{Quillen} \& {Garnett}}{{Quillen} \&
  {Garnett}}{2000}]{Quillen00}
{Quillen} A.~C.,  {Garnett} D.~R.,  2000, ArXiv Astrophysics e-prints, \href
  {http://adsabs.harvard.edu/abs/2000astro.ph..4210Q} {}

\bibitem[\protect\citeauthoryear{{Robin}, {Reyl{\'e}}, {Derri{\`e}re}  \&
  {Picaud}}{{Robin} et~al.}{2003}]{Bes}
{Robin} A.~C.,  {Reyl{\'e}} C.,  {Derri{\`e}re} S.,   {Picaud} S.,  2003,
  \mn@doi [A\&A] {10.1051/0004-6361:20031117}, \href
  {http://adsabs.harvard.edu/abs/2003A%26A...409..523R} {409, 523}

\bibitem[\protect\citeauthoryear{{Sackett}}{{Sackett}}{1997}]{Sackett:97}
{Sackett} P.~D.,  1997, ApJ, \href
  {http://adsabs.harvard.edu/abs/1997ApJ...483..103S} {483, 103}

\bibitem[\protect\citeauthoryear{{Soderblom}}{{Soderblom}}{2010}]{Soderblom2010}
{Soderblom} D.~R.,  2010, \mn@doi [ARA\&A]
  {10.1146/annurev-astro-081309-130806}, \href
  {http://adsabs.harvard.edu/abs/2010ARA%26A..48..581S} {48, 581}

\bibitem[\protect\citeauthoryear{{Tully} \& {Fouqu\'e}}{{Tully} \&
  {Fouqu\'e}}{1985}]{Tully:1985}
{Tully} R.~B.,  {Fouqu\'e} P.,  1985, \mn@doi [ApJS] {10.1086/191029}, \href
  {http://adsabs.harvard.edu/abs/1985ApJS...58...67T} {58, 67}

\bibitem[\protect\citeauthoryear{{Wielen}}{{Wielen}}{1977}]{Wielen77}
{Wielen} R.,  1977, A\&A, \href
  {http://adsabs.harvard.edu/abs/1977A%26A....60..263W} {60, 263}

\bibitem[\protect\citeauthoryear{{Yoachim} \& {Dalcanton}}{{Yoachim} \&
  {Dalcanton}}{2006}]{YandD}
{Yoachim} P.,  {Dalcanton} J.~J.,  2006, \mn@doi [AJ] {10.1086/497970}, \href
  {http://adsabs.harvard.edu/abs/2006AJ....131..226Y} {131, 226}

\bibitem[\protect\citeauthoryear{{Zacharias}, {Finch}, {Girard}, {Henden},
  {Bartlett}, {Monet}  \& {Zacharias}}{{Zacharias} et~al.}{2013}]{UCAC4}
{Zacharias} N.,  {Finch} C.~T.,  {Girard} T.~M.,  {Henden} A.,  {Bartlett}
  J.~L.,  {Monet} D.~G.,   {Zacharias} M.~I.,  2013, \mn@doi [AJ]
  {10.1088/0004-6256/145/2/44}, \href
  {http://adsabs.harvard.edu/abs/2013AJ....145...44Z} {145, 44}

\bibitem[\protect\citeauthoryear{{de Grijs}, {Peletier}  \& {van der
  Kruit}}{{de Grijs} et~al.}{1997}]{de-Grijs:97}
{de Grijs} R.,  {Peletier} R.~F.,   {van der Kruit} P.~C.,  1997, A\&A, \href
  {http://adsabs.harvard.edu/abs/1997A%26A...327..966D} {327, 966}

\bibitem[\protect\citeauthoryear{{van Albada}, {Bahcall}, {Begeman}  \&
  {Sancisi}}{{van Albada} et~al.}{1985}]{vanAlbada}
{van Albada} T.~S.,  {Bahcall} J.~N.,  {Begeman} K.,   {Sancisi} R.,  1985,
  \mn@doi [ApJ] {10.1086/163375}, \href
  {http://adsabs.harvard.edu/abs/1985ApJ...295..305V} {295, 305}

\bibitem[\protect\citeauthoryear{{van der Kruit} \& {Freeman}}{{van der Kruit}
  \& {Freeman}}{1984}]{VanFree}
{van der Kruit} P.~C.,  {Freeman} K.~C.,  1984, \mn@doi [ApJ] {10.1086/161769},
  \href {http://adsabs.harvard.edu/abs/1984ApJ...278...81V} {278, 81}

\bibitem[\protect\citeauthoryear{{van der Kruit} \& {Freeman}}{{van der Kruit}
  \& {Freeman}}{2011}]{vdKF2011}
{van der Kruit} P.~C.,  {Freeman} K.~C.,  2011, \mn@doi [ARA\&A]
  {10.1146/annurev-astro-083109-153241}, \href
  {http://adsabs.harvard.edu/abs/2011ARA%26A..49..301V} {49, 301}

\makeatother
\end{thebibliography}
\vspace*{1cm}

\noindent {\bf APPENDIX: Contribution of different parts of the CMD to the integrated spectra of the disc}

The Galactic disc near the sun is a composite population, with stars covering the whole range of age from very young to about 10 Gyr. The stellar population is similarly composite in other star-forming galaxies; the details depend on the local star formation history. As a guide to the relative contributions to the integrated light in external galaxies, from young and old giants, bright main sequence stars, and the main sequence below the old turnoff, we use the IAC-STAR synthetic CMD computation code to evaluate the contributions from the different parts of the CMD to the integrated light of the disc for a star formation history like that described in section 1 (see Fig. 1 and related text). The region of the spectrum that is mostly used for integrated light spectroscopy of the disc is around $5200$\AA\ near the Mg b band. We are therefore interested in the relative contributions to the integrated light at the $V$-band.

Fig. 1 illustrates the regions of the CMD that are populated by younger and older stars. We partition the CMD
somewhat arbitrarily into 3 regions: the bright main sequence ($M_V \leq 4, B - V \leq 0.7$), the lower main sequence ($M_V > 4$) and the red giants ($M_V \leq 4, B - V > 0.7$). For each of these 3 regions of the CMD, we constructed the $M_V$ luminosity function for the stars in the IAC simulated catalogue. For the adopted star formation history ($ \propto \exp(-t/\beta$ with $\beta = 20$ Gyr), the contributions from the three regions to the $V$-band light are: bright main sequence $37\%$, lower main sequence 10\% and red giants 53\%. The young giants (age $\leq 2$ Gyr) contributed 16\% of the total light, and the older giants contributed 37\% of the total light. These relative contributions depend on the adopted star formation history. For a more nearly constant star formation history, with $\beta = 60$ Gyr, the younger and older giants contribute 16\% and 32\% respectively to the integrated light.

Although the spectra of stars around the main sequence turnoff do contribute to the equivalent widths of the absorption lines in the Mg b region, the red giants are the main contributors to the absorption lines in the integrated spectra from which the velocity dispersions are measured. This is why we chose to concentrate on the kinematics of the red giants in the solar neighbourhood in the present paper. Comparison of the integrated spectra of galactic discs with the spectra of individual giant stars shows that the lines of the integrated spectra are diluted by the contribution of continuum light from the hotter main sequence stars. 

The young giants contribute 
30\% to 35\% of the total light from the giants. The point of this paper is that the contribution from these stars needs to be taken into account when measuring the velocity dispersion from the integrated light of external galaxies, to avoid underestimating the surface mass density of the galactic disc.

\begin{figure} 
\includegraphics[width=0.5\textwidth]{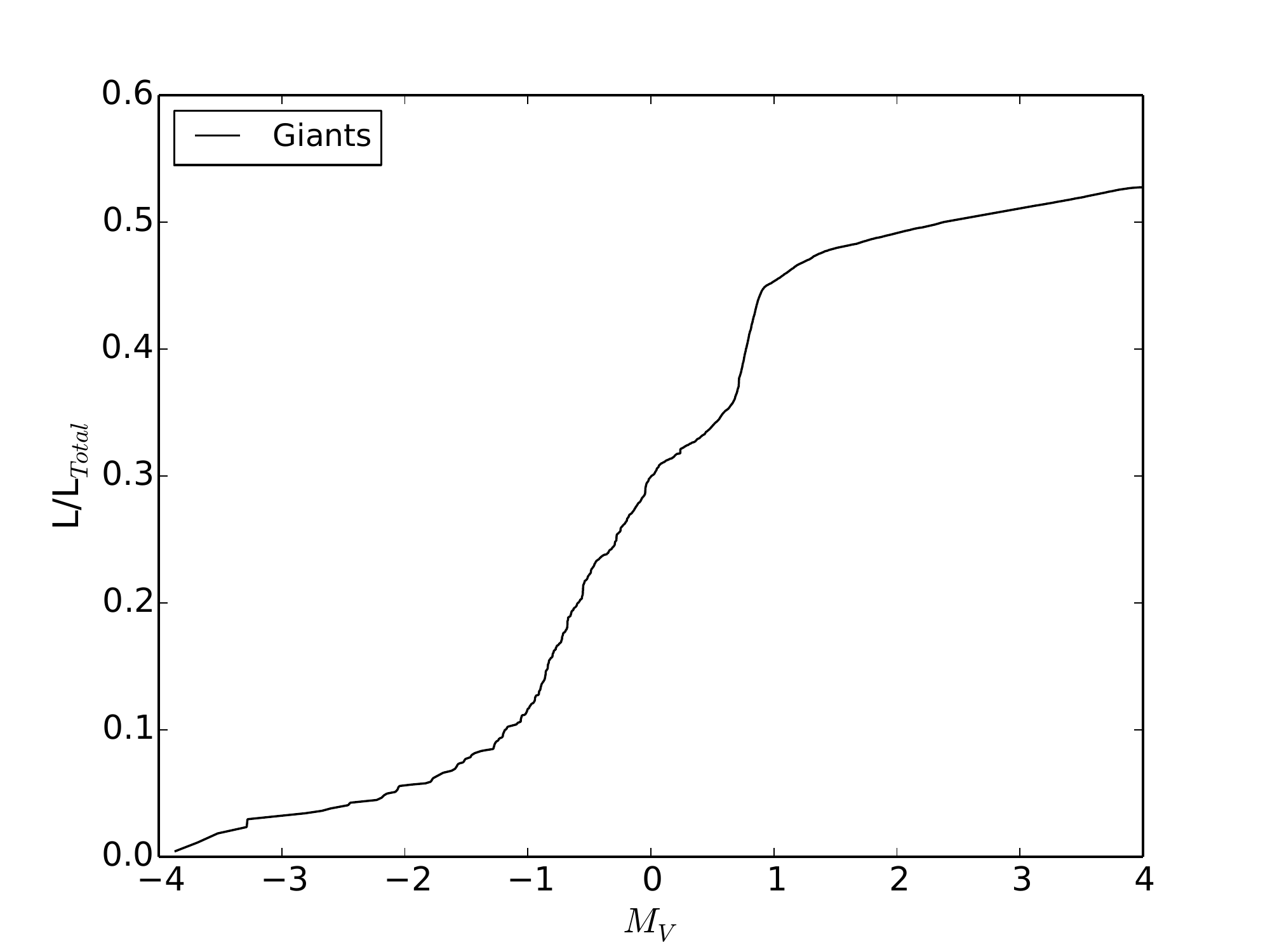} 
\caption{Cumulative histogram of the integrated light of red giants derived from the IAC-STAR simulation. 
$L_{total}$ is the total light from all stars. The giants provide about $53$\% of the total surface brightness of the disc, for the adopted star formation history $\beta = 20$ Gyr. }
\end{figure}
\end{document}